\documentstyle[preprint,epsfig,eqsecnum,aps]{revtex}
\newcommand{\vslash}{\mbox{$\not{\hspace{-1.03mm}v}$}}        
\newcommand{\bea}{\begin{eqnarray}}
\newcommand{\eea}{\end{eqnarray}}
\newcommand{\beq}{\begin{equation}}
\newcommand{\eeq}{\end{equation}}
\newcommand{\bay}{\begin{array}}
\newcommand{\eay}{\end{array}}
\begin{document}
\preprint{\parbox{6cm}{\flushright CNLS 97/1457\\TECHNION-PH 97-01}}
\title{Predictions for s-Wave and p-Wave Heavy Baryons\\ 
from Sum Rules and Constituent Quark Model\\
(I): Strong Interactions}
\author{Dan Pirjol}
\address{Department of Physics, Technion - Israel Institute of Technology,
32000 Haifa, Israel}
\author{Tung-Mow Yan}
\address{Floyd R. Newman Laboratory of Nuclear Studies, Cornell University,
Ithaca, New York 14853}
\date{\today}
\maketitle
\begin{abstract}
We study the strong interactions of the $L=1$ orbitally excited
baryons with one heavy quark in the framework of the Heavy Hadron
Chiral Perturbation Theory. To leading order in the heavy mass
expansion, the interaction Lagrangian
describing the couplings of these states among themselves and with
the ground state
heavy baryons contains 46 unknown couplings. We derive sum rules
analogous to the Adler-Weisberger sum rule which
constrain these couplings and relate them to the couplings of
the s-wave heavy baryons. Using a spin 3/2 baryon as a target,
we find a sum rule expressing
the deviation from the quark model prediction for pion couplings
to s-wave states in terms of couplings of the p-wave states.
In the constituent quark model these
couplings are related and can be expressed in terms of
only two reduced matrix elements. Using recent CLEO data on 
$\Sigma_c^{*}$ and $\Lambda_{c1}^+$ strong decays, we determine
some of the unknown couplings in the chiral Lagrangian and the two
quark model reduced matrix elements. Specific predictions are 
made for the decay properties of all $L=1$ charmed baryons.
\end{abstract}
\narrowtext
\section{Introduction}
Baryons containing one heavy quark offer an important testing ground
for the ideas and predictions of heavy quark spin-flavor SU(4) and light
flavor SU(3) symmetries. These symmetries become manifest in QCD in the
limits of infinite heavy quark masses $m_b,m_c\to\infty$ and identical
light quark masses $m_u=m_d=m_s$. Although implications of these symmetries 
for the spectroscopy and decay properties of the heavy baryons are
well-known (for a review see e.g. \cite{KKP}), so far very few
predictions, if any, can be compared with experiments due to lack of
both data and theoretical knowledge of the unknown parameters.

This is the first of a series of papers in which we will study the
properties of the excited heavy baryons,
focusing on the first orbital excitations, the p-wave baryons with 
one heavy quark. In a sequel we will consider the radiative decays
of these states.

The spectroscopy of these states is reviewed in
Section II in the language of the constituent quark model.
The constraints imposed by heavy
quark symmetry on the possible structure of these couplings can be
automatically incorporated by describing them in the framework of heavy 
hadron chiral perturbation theory \cite{hhcpt,yan}. The resulting chiral
Lagrangian is presented in Section III. We include all possible
strong interaction couplings among and between s-wave and p-wave
baryons to leading order in $1/m_Q$ and chiral expansion. There are
a total of 46 independent coupling constants up to and including
D-wave interactions, which in principle, have to be extracted
from experiment. Recent data from FNAL \cite{LC0,LC2} and CLEO \cite{LC1}
make it possible to test and constrain the parameters of the theory.

We derive in Section IV model-independent sum rules which constrain
these couplings and relate them to properties of the lowest-lying baryons. 
For the strong decay amplitudes these sum rules can be derived in 
analogy with
the Adler-Weisberger (AW) sum rule familiar from current algebra.
With a spin 3/2 baryon as a target, the two spin projections 3/2
and 1/2 along the incident pion's momentum give rise to two sum rules.
One of these can be used to
parametrize the deviation from the
quark model relation among the two pion couplings $g_1,g_2$ to
the s-wave heavy baryons, expressing it in terms of the pion
couplings of the p-wave baryons. 

In Section V we derive predictions for the strong couplings of the p-wave 
baryons in the constituent quark model. Many of these coupling
constants can be computed
in the quark model whereas others can be related in a simple way.
In fact all but two of the 46 coupling constants are determined
in the quark model. Furthermore, these two couplings are constrained
to satisfy an AW sum rule.

In Section VI we discuss a few phenomenological applications
of our results. We extract one of the pion couplings to the s-wave
baryons $g_2$ from recent CLEO measurements on the $\Sigma_c^{*}$
width. The extracted value for $g_2$ is consistent with the quark
model prediction.
This is used in turn to determine the S-wave and D-wave 
couplings of two p-wave
charmed baryons $\Lambda_{c1}^+$ from their two-pion widths.
Taken together with the quark model relations
in Section V, these couplings can be used  to estimate
the strong couplings of all p-wave charmed baryons. A few
specific predictions are presented for some decay modes of
these states. We conclude with some comments in Section VII.

\section{Spectroscopy of heavy baryons}

The heavy baryons fall into the ${\bf\bar 3}$ and {\bf 6} representations
of flavor SU(3), into which the product {\bf 3}$\otimes${\bf 3} =
${\bf\bar 3}\oplus {\bf 6}$ is decomposed, corresponding to the two
light quarks in the baryon. The lowest-lying states transform as an
${\bf\bar 3}$ and can be represented either as an antisymmetric matrix
$B_{\bar 3}$ \cite{yan} or as a vector $T$ \cite{cho}
\bea\label{1}
T^i = \frac{1+\vslash}{2}\left(\Xi_c^0\quad -\Xi_c^+\quad \Lambda_c^+
\right)_i  = \frac12 \epsilon_{ijk}(B_{\bar 3})_{jk}\,.
\eea
We have taken as heavy quark a charm quark.
This multiplet contains an isospin doublet $(\Xi_c^0\quad -\Xi_c^+)$ and
a singlet $\Lambda_c^+$.
In the heavy quark limit, the angular momentum and parity of the light 
constituents in a heavy baryon become good quantum numbers and the multiplet
(\ref{1}) has $s_\ell^{\pi_\ell}=0^+$. 

Above this multiplet lie other s-wave states with $s_\ell^{\pi_\ell}=1^+$
which transform as a {\bf 6} under light SU(3). When combining the spin 1
of the light degrees of freedom with the heavy quark spin 1/2, one almost
degenerate doublet is obtained, with total spins $J=1/2,3/2$. Both these states
can be grouped together into one superfield as \cite{cho}
\bea\label{2}
S_\mu^{ij} = \frac{1}{\sqrt3}(\gamma_\mu+v_\mu)\gamma_5\frac{1+\vslash}{2}
B_6^{ij} + \frac{1+\vslash}{2}
B^{*ij}_{6\mu}\,.
\eea
The matrices $B_6$ and $B^*_{6\mu}$ are defined in \cite{yan}
\bea\label{3}
(B_6)_{ij} = 
\left( \begin{array}{ccc}
\Sigma_c^{++} & \frac{1}{\sqrt2}\Sigma_c^+ & \frac{1}{\sqrt2}\Xi_c^{+'} \\
\frac{1}{\sqrt2}\Sigma_c^+ & \Sigma_c^0 & \frac{1}{\sqrt2}\Xi_c^{0'} \\
\frac{1}{\sqrt2}\Xi_c^{+'} & \frac{1}{\sqrt2}\Xi_c^{0'} & \Omega_c^0
\end{array} \right)_{ij}
\eea
and analogously for the sextet of spin 3/2 Rarita-Schwinger baryon fields
$B^*_{6\mu}$.

\begin{center}
\begin{tabular}{|c|c|c|c|}
\hline
\qquad\qquad\qquad & \quad SU(3)\quad\quad & \quad S\quad\quad & 
\quad $s_\ell^{\pi_\ell}$\quad\quad \\
\hline
$\Lambda_{c1}(\frac12,\frac32)$ & $\bar 3$ & 0 & $1^-$ \\
$\Sigma_{c0}(\frac12)$ & $6$ & 1 & $0^-$ \\
$\Sigma_{c1}(\frac12,\frac32)$ & $6$ & 1 & $1^-$ \\
$\Sigma_{c2}(\frac32,\frac52)$ & $6$ & 1 & $2^-$ \\
\hline
$\Sigma'_{c1}(\frac12,\frac32)$ & $6$ & 0 & $1^-$ \\
$\Lambda'_{c0}(\frac12)$ & $\bar 3$ & 1 & $0^-$ \\
$\Lambda'_{c1}(\frac12,\frac32)$ & $\bar 3$ & 1 & $1^-$ \\
$\Lambda'_{c2}(\frac32,\frac52)$ & $\bar 3$ & 1 & $2^-$ \\
\hline
\end{tabular}
\end{center}
\begin{quote}
{\bf Table 1.}
The p-wave charmed baryons and their quantum numbers. 
$S$ (the total spin of the two light
quarks) is a good quantum number only in the constituent quark model. 
In the quark model, 
the first (last) four multiplets have even (odd) orbital wavefunctions 
under a permutation
of the two light quarks.\\
\end{quote}

The spectroscopy of the p-wave heavy baryons is more complex. There are
altogether eight heavy quark symmetry multiplets of p-wave baryons, represented
in Table 1 together with their quantum numbers\footnote{The terminology adopted
here is particularly suggestive, with the subscript labeling the angular momentum
of the light degrees of freedom.}.
They can be classified into two distinct groups, corresponding in the constituent
quark model to symmetric and antisymmetric orbital wavefunctions, respectively,
under a permutation
of the two light quarks \cite{CIK,IWY}. We will refer to them as symmetric
and antisymmetric states. Potential models \cite{CIK,CI} indicate that
the former lie about 150 MeV below the latter.

The lowest-lying p-wave states
arise from combining the heavy quark spin with light constituents in a
$s_\ell^{\pi_\ell}=1^-$ symmetric state. The corresponding heavy baryon states
have spin and parity $J^P=1/2^-,3/2^-$. The $I=0$ members of these multiplets
have been observed experimentally \cite{LC-1,LC0,LC1,LC2} and are known as
$\Lambda_{c_1}(\frac12,\frac32)$. Their fields can be combined again into a
superfield as \cite{cho2}
\bea\label{4}
R_\mu^i = \frac{1}{\sqrt3}(\gamma_\mu+v_\mu)\gamma_5 R^i + R^{*i}_\mu
\eea
with 
\bea\label{5}
R_i = \frac{1+\vslash}{2}\left(\Xi_{c_1}^0\quad -\Xi_{c_1}^+\quad
\Lambda_{c_1}^+\right)_i\,,\qquad
R_\mu^{*i} = \frac{1+\vslash}{2}\left(\Xi_{c_1\mu}^{*0}\quad
 -\Xi_{c_1\mu}^{*+}\quad \Lambda_{c_1\mu}^{*+}\right)_i\,. 
\eea

Above these states lie three other p-wave {\bf 6} symmetric multiplets
with quantum numbers of the light degrees of freedom $s_\ell^{\pi_\ell}
=0^-,1^-,2^-$. Their $I=1$ members will be denoted as
$\Sigma_{c0}(\frac12),
\Sigma_{c1}(\frac12,\frac32)$ and $\Sigma_{c2}(\frac32,\frac52)$.
The $s_\ell^{\pi_\ell}=0^-$ multiplet will be represented as a symmetric
matrix $(U)_{ij}$ defined as in (\ref{3}) and the $s_\ell^{\pi_\ell}=1^-$
multiplet will be represented as a superfield similar to (\ref{4}) but with
a symmetric matrix $V_\mu^{ij}$
\bea\label{5.1}
V_\mu^{ij} = \frac{1}{\sqrt3}(\gamma_\mu+v_\mu)\gamma_5 V^{ij} + V^{*ij}_\mu
\eea
The superfield corresponding to the $s_\ell^{\pi_\ell}=2^-$ baryons is
constructed as \cite{Fa}
\bea
X_{\mu\nu}^{ij} = X_{\mu\nu}^{*ij} +
\frac{1}{\sqrt{10}}\left\{(\gamma_\mu+v_\mu)\gamma_5 g_{\nu\alpha} +
(\gamma_\nu+v_\nu)\gamma_5 g_{\mu\alpha}\right\}X_\alpha^{ij}
\eea
with $X_{\mu\nu}^{*ij}$ a spin-5/2 Rarita-Schwinger field and $X_\alpha^{ij}$
its spin-3/2 heavy quark symmetry partner.

The antisymmetric p-wave states are constructed in complete analogy to the
symmetric ones. There is a sextet $\Sigma'_{c1}(\frac12,\frac32)$ with quantum
numbers
$s_\ell^{\pi_\ell}=1^-$, which will be represented again by a superfield
$R^{'ij}_\mu$ constructed in analogy to (\ref{2}). In addition to this,
there are three antitriplets, whose $I=0$ members are denoted by
$\Lambda_{c0}^{'+},\Lambda_{c1}^{'+},\Lambda_{c2}^{'+}$. Their superfields
will be denoted as $U'_i,V^{'i}_\mu,X^{'i}_{\mu\nu}$.

\section{Strong couplings of the heavy baryons}

The couplings of the heavy baryons to the Goldstone bosons are described 
most compactly when expressed in terms of their superfields
(\ref{2},\ref{4},\ref{5.1}). The leading terms describe P-wave
couplings among the s-wave baryons and S-wave couplings between the
s-wave and p-wave baryons
\bea\label{6}
{\cal L}_{int} &=& 
\frac32 ig_1 \epsilon_{\mu\nu\sigma\lambda}\mbox{tr}(\bar S^\mu v^\nu
A^\sigma S^\lambda)
- \sqrt3 g_2 \mbox{tr}\left(\bar B_{\bar 3}A^\mu S_\mu + \bar S^\mu A_\mu
B_{\bar 3}\right)\\
& &+ h_2\left\{ \epsilon_{ijk}\bar R^i_\mu v_\nu A^\nu_{jl}S_\mu^{kl}+
\epsilon_{ijk}\bar S^{kl}_\mu v_\nu A^\nu_{lj} R^i_\mu\right\} + h_3
\mbox{tr}\left(\bar B_{\bar 3}v_\mu A^\mu U + \bar U v^\mu A_\mu
B_{\bar 3}\right)\nonumber\\
& &+ h_4\mbox{tr}\left\{ \bar V_\mu v_\nu A^\nu S_\mu +
\bar S_\mu v_\nu A^\nu V_\mu\right\}\nonumber\\
& &+ h_5\mbox{tr}\left( \bar R'_\mu v_\nu A^\nu S^\mu +
\bar S^\mu v_\nu A^\nu R'_\mu\right) +
h_6\left( \bar T_i v_\nu A^\nu_{ji}U'_j +
\bar U'_i v_\nu A^\nu_{ji}T_j\right)\nonumber\\
& &+h_7\left\{ \epsilon_{ijk}\bar V^{'i}_\mu v_\nu A^\nu_{jl}S^\mu_{kl}
+ \epsilon_{ijk}\bar S^\mu_{kl} v_\nu A^\nu_{lj}V^{'i}_\mu\right\}\nonumber\,.
\eea
The Goldstone bosons couple to the matter fields through the nonlinear
axial field $A_\mu$ defined as
\bea
A_\mu = \frac{i}{2}\left( \xi^\dagger\partial_\mu\xi 
- \xi\partial_\mu\xi^\dagger \right)\,,
\eea
with $\xi=\exp(iM/f_\pi)$, $M=\frac{1}{\sqrt2}\pi^a\lambda^a$ and $f_\pi=132$ MeV
\cite{hhcpt,yan}.

The couplings $g_1,g_2$ are defined as in \cite{yan}\footnote{The
couplings in \cite{cho} are related to the ones in (\ref{6})
by $(g_2)_{Cho}=3/2g_1$ and $(g_3)_{Cho}=-\sqrt3 g_2$.}
and the coupling of the ${\bf \bar 3}$ p-wave baryons $h_2$ is chosen as in \cite{cho2}. 
We introduced new constants $h_3-h_7$ describing all the S-wave pion couplings
of the p-wave to the s-wave baryons which are allowed by heavy quark symmetry.

The D-wave couplings of the p-wave baryons to s-wave baryons are described by
dimension-5 terms in the effective Lagrangian
\bea\label{6D}
{\cal L}_{D} &=& 
ih_8\epsilon_{ijk} \bar S_\mu^{kl}\left( {\cal D}_\mu A_\nu + {\cal D}_\nu A_\mu
+ \frac23 g_{\mu\nu}(v\cdot {\cal D})(v\cdot A)\right)_{lj}R_\nu^i\\
&+& ih_9 \mbox{tr }\left\{\bar S_\mu \left( {\cal D}_\mu A_\nu + {\cal D}_\nu A_\mu
+ \frac23 g_{\mu\nu}(v\cdot {\cal D})(v\cdot A)\right) V_\nu\right\}\nonumber\\
&+& 
ih_{10} \epsilon_{ijk} \bar T_i\left( {\cal D}_\mu A_\nu + {\cal D}_\nu A_\mu\right)_{jl}
X_{kl}^{\mu\nu} + h_{11}\epsilon_{\mu\nu\sigma\lambda}\mbox{tr }\left\{
\bar S_{\mu}\left( {\cal D}_\nu A_\alpha + {\cal D}_\alpha A_\nu\right)X_{\alpha\sigma}
\right\}v_\lambda\nonumber\\
&+& ih_{12}\mbox{tr }\left\{
\bar S_\mu \left( {\cal D}_\mu A_\nu + {\cal D}_\nu A_\mu
+ \frac23 g_{\mu\nu}(v\cdot {\cal D})(v\cdot A)\right) R'_\nu\right\}\nonumber\\
&+& ih_{13}\epsilon_{ijk} \bar S_\mu^{kl}
\left( {\cal D}_\mu A_\nu + {\cal D}_\nu A_\mu
+ \frac23 g_{\mu\nu}(v\cdot {\cal D})(v\cdot A)\right)_{lj} V_\nu^{'i}\nonumber\\
&+& ih_{14}\bar T_i\left( {\cal D}_\mu A_\nu + {\cal D}_\nu A_\mu\right)_{ji}
X_{\mu\nu}^{'j} + h_{15}\epsilon_{\mu\nu\sigma\lambda}\epsilon_{ijk}
\bar S_{\mu}^{kl}\left( {\cal D}_\nu A_\alpha + {\cal D}_\alpha A_\nu\right)_{lj}
X_{\alpha\sigma}^{'i} v_\lambda\nonumber\,.
\eea
The covariant derivative of the axial field $A_\mu$ is defined as
${\cal D}_\mu A_\nu = \partial_\mu A_\nu + [V_\mu\,,A_\nu]$ with
\bea
V_\mu = \frac{1}{2}\left( \xi^\dagger\partial_\mu\xi 
+ \xi\partial_\mu\xi^\dagger \right)\,,
\eea
and satisfies the relation ${\cal D}_\mu A_\nu - {\cal D}_\nu A_\mu = 0$.
The structure ${\cal D}_\mu A_\nu + {\cal D}_\nu A_\mu
+ \frac23 g_{\mu\nu}(v\cdot {\cal D})(v\cdot A)$ appearing in the $\bar SR$
and $\bar SV$ couplings projects out a pure D-wave.

  In addition to the couplings described by the Lagrangians (\ref{6},\ref{6D}),
the p-wave baryons can couple also among themselves and to the Goldstone bosons.
The most general Lagrangian allowed by heavy quark symmetry describing
the P-wave couplings of the symmetric p-wave states has the form
\bea\label{Lsymm}
{\cal L}_{pp\pi} &=& 
if_1\epsilon^{\mu\nu\sigma\lambda}\bar R^i_\mu v_\nu A_\sigma^{ji} R^j_\lambda
+ if_2\epsilon^{\mu\nu\sigma\lambda}
\mbox{tr }(\bar V_\mu v_\nu A_\sigma V_\lambda) +
if_3 \epsilon^{\mu\nu\sigma\lambda}\mbox{tr }(\bar X_{\mu\alpha}
v_\nu A_\sigma X_{\alpha\lambda})\\
&+& 
f_4\epsilon_{ijk} \bar R^{i\mu} A_\mu^{jl} U^{kl}
+ if_5\epsilon^{\mu\nu\sigma\lambda} \epsilon_{ijk} \bar R^{i\mu}
v_\nu A_\sigma^{jl} V_\lambda^{kl} +
f_6\epsilon_{ijk} \left(\bar R^{i\mu} A^{\nu jl}+\bar R^{i\nu} A^{\mu jl}\right)
X_{\mu\nu}^{kl}\nonumber\\
&+& f_7\mbox{tr }(\bar U A^\mu V_\mu)
+ f_8 \mbox{tr }[(\bar V_\mu A_\nu + \bar V_\nu A_\mu)X_{\mu\nu}]
+ \mbox{h.c.}\nonumber\,.
\eea
We note that the Goldstone bosons do not couple to the field $U$ alone,
as the only possible coupling tr $(\bar Uv\cdot AU)$ does not conserve
parity.

The couplings of the antisymmetric p-wave states to the Goldstone bosons
are described by a Lagrangian similar to (\ref{Lsymm})
\bea\label{Lasymm}
{\cal L}_{p'p'\pi} &=& 
if'_1\epsilon^{\mu\nu\sigma\lambda}
\mbox{tr }(\bar R'_\mu v_\nu A_\sigma R'_\lambda)
+ if'_2\epsilon^{\mu\nu\sigma\lambda}
\bar V^{'i}_\mu v_\nu A^{ji}_\sigma V^{'j}_\lambda +
if'_3 \epsilon^{\mu\nu\sigma\lambda} \bar X^{'i}_{\mu\alpha}
v_\nu A^{ji}_\sigma X^{'j}_{\alpha\lambda}\\
&+& 
f'_4\epsilon_{ijk} \bar R^{'kl\mu} A_\mu^{lj} U^{'i}
+ if'_5\epsilon^{\mu\nu\sigma\lambda} \epsilon_{ijk} \bar R^{'kl\mu}
v_\nu A_\sigma^{lj} V_\lambda^{'i} +
f'_6\epsilon_{ijk} \left(\bar R^{'kl\mu} A^{\nu lj}+\bar R^{'kl\nu} A^{\mu lj}
\right) X_{\mu\nu}^{'i}\nonumber\\
&+& f'_7 \bar U^{'i} A^{ji\mu} V^{'j}_\mu
+ f'_8 (\bar V^{'i}_\mu A_\nu^{ji} + \bar V^{'i}_\nu A^{ji}_\mu)X^{'j}_{\mu\nu}
+ \mbox{h.c.}\nonumber\,.
\eea

Finally, the couplings of the symmetric to antisymmetric p-wave states are
given by a Lagrangian containing fourteen additional couplings
\bea\label{Lstoas}
{\cal L}_{pp'\pi} &=& 
if_1''\epsilon_{\mu\nu\sigma\lambda}\epsilon_{ijk}
\bar R_\mu^i v_\nu A_\sigma^{jl}R^{'kl}_\lambda +
f_2''\mbox{tr}(\bar UA_\mu R'_\mu) +
if_3''\epsilon_{\mu\nu\sigma\lambda}\mbox{tr}(\bar V_\mu v_\nu A_\sigma R'_\lambda)\\
&+& f_4''\mbox{tr}[\bar X_{\mu\nu}(A_\mu R'_\nu + A_\nu R'_\mu)] +
if_5''\epsilon_{\mu\nu\sigma\lambda} \bar R_\mu^i v_\nu A_\sigma^{ji} U^{'j}_\lambda
+ f_6''\epsilon_{ijk} \bar V^{kl}A_\mu^{jl} U^{'i}_\mu\nonumber\\
&+& f_7''\epsilon_{ijk} \bar X_{\mu\nu}^{kl}(A_\nu^{jl} U^{'i}_\mu +
A_\mu^{jl} U^{'i}_\nu)) +
if_8''\epsilon_{\mu\nu\sigma\lambda} \bar R^i_\mu v_\nu A_\sigma^{ji}V^{'j}_\lambda
+ f_9''\epsilon_{ijk} \bar U^{kl}A_\mu^{jl} V^{'i}_\mu\nonumber\\
&+& if_{10}''\epsilon_{\mu\nu\sigma\lambda}\epsilon_{ijk}
\bar V_\mu^{kl} v_\nu A_\sigma^{jl} V^{'i}_\lambda +
f_{11}''\epsilon_{ijk} \bar X_{\mu\nu}^{kl}
(A_\nu^{jl}V_\mu^{'i}+A_\mu^{jl}V_\nu^{'i}) +
f_{12}''(\bar R_\mu^i A_\nu^{ji}+\bar R_\nu^i A_\mu^{ji})X^{'j}_{\mu\nu}\nonumber\\
&+& f_{13}''\epsilon_{ijk}(\bar V_\mu^{kl} A_\nu^{jl}+\bar V_\nu^{kl} A_\mu^{jl})
X^{'i}_{\mu\nu} +
if_{14}''\epsilon_{ijk}\epsilon_{\mu\alpha\beta\lambda} \bar X_{\mu\nu}^{kl}
v_\alpha A_\beta^{jl} X^{'i}_{\nu\lambda}\nonumber\,.
\eea
We neglected in (\ref{Lsymm},\ref{Lasymm},\ref{Lstoas}) interaction terms
describing F-wave couplings, as they are expected to be highly suppressed
on dimensional grounds.

The Lagrangian (\ref{6})
gives the following typical decay widths
\bea\label{7}
\Gamma(\Sigma_c^{++}\to\pi^+\Lambda_c^+) &=& \frac{g_2^2}{2\pi f_\pi^2}
\frac{M_{\Lambda_c^+}}{M_{\Sigma_c^{++}}}|\vec p_\pi\,|^3\,,\qquad
\Gamma(\Sigma_c^{++*}\to\pi^+\Sigma_c^+) = \frac{g_1^2}{16\pi f_\pi^2}
\frac{M_{\Sigma_c^+}}{M_{\Sigma_c^{++*}}}|\vec p_\pi\,|^3\,,\\
\Gamma(\Lambda_{c_1}^+(\frac12)\to\pi^+\Sigma_c^0) &=& \frac{h_2^2}{2\pi f_\pi^2}
\frac{M_{\Sigma_c^0}}{M_{\Lambda_{c1}^+}}E_\pi^2|\vec p_\pi\,|\,,\qquad
\Gamma(\Sigma_{c0}^{++}(\frac12)\to\pi^+\Lambda_c^+) = \frac{h_3^2}{2\pi f_\pi^2}
\frac{M_{\Lambda_c^+}}{M_{\Sigma_{c0}^{++}}}E_\pi^2|\vec p_\pi\,|\,,\nonumber\\
\Gamma(\Sigma_{c1}^{++}(\frac12)\to\pi^+\Sigma_c^+) &=& \frac{h_4^2}{4\pi f_\pi^2}
\frac{M_{\Sigma_c^+}}{M_{\Sigma_{c1}^{++}}}E_\pi^2|\vec p_\pi\,|\,,\qquad
\Gamma(\Sigma_{c1}^{'++}(\frac12)\to\pi^+\Sigma_c^+) = \frac{h_5^2}{4\pi f_\pi^2}
\frac{M_{\Sigma_c^+}}{M_{\Sigma_{c1}^{'++}}}E_\pi^2|\vec p_\pi\,|\,,\nonumber\\
\Gamma(\Xi_{c0}^{'0}(\frac12)\to\pi^-\Xi_c^+) &=& \frac{h_6^2}{2\pi f_\pi^2}
\frac{M_{\Xi_c^+}}{M_{\Xi_{c0}^{'0}}}E_\pi^2|\vec p_\pi\,|\,,\qquad
\Gamma(\Lambda_{c1}^{'+}(\frac12)\to\pi^+\Sigma_c^0) = \frac{h_7^2}{2\pi f_\pi^2}
\frac{M_{\Sigma_c^0}}{M_{\Lambda_{c1}^{'+}}}E_\pi^2|\vec p_\pi\,|\,.\nonumber
\eea

\section{Adler-Weisberger sum rules for heavy baryons}
\subsection{Narrow width sum rules}

One can derive an analog of the Adler-Weisberger sum rule involving the
coupling $g_2$  by considering a dispersion relation for pion scattering
on an s-wave ${\bf \bar 3}$ baryon (similar sum rules have been discussed 
in \cite{DP,P2,Be,CP,MB} for the heavy meson case).
One possible derivation, which will
prove most convenient in the following, is based on the use of the forward
(spin-averaged) matrix element of the retarded commutator\footnote{The derivation
presented here is a slightly modified version of the one given in \cite{FFR}.}
\bea\label{8}
F^{ab}_{ji}(\nu) &=& i\int\mbox{d}^4x\, e^{iq\cdot x}\theta(x_0)
\langle P_j|[D^a(x),\, D^b(0)]|P_i\rangle\\
 &=& F^{(+)}(\nu)\frac12 \{\tau^a,\tau^b\}_{ji} +
F^{(-)}(\nu)\frac12 [\tau^a,\tau^b]_{ji}\nonumber
\eea
where $D^a=\partial^\mu(\bar q\gamma_\mu\frac{\tau^a}{2} q)$ and
$\nu=q^0$ is the pion energy in the baryon rest frame. 
Usual manipulations with current density commutators give the relation \cite{FFR}
\bea\label{8a}
\frac{\partial}{\partial\nu}F^{(-)}(\nu)\vert_{q=0} = I_3
\eea
with $I_3$ the isospin of the target. The states in (\ref{8}) are normalised
according to $\langle P_i|P_j\rangle = (E_i/m)(2\pi)^3\delta(\vec P_i-\vec P_j)$.

Inserting a complete set of states in (\ref{8}) gives
\bea\label{8b}
F^{ab}_{ji}(\nu) &=& (2\pi)^3\sum_\Gamma\int\mbox{d}\mu(\Gamma)
\frac{\langle P_j|D^{a}(0)|\Gamma\rangle\langle\Gamma |D^b(0)|P_i\rangle}
{-\nu-E+E_\Gamma-i\epsilon}\delta(\vec q+\vec P-\vec P_\Gamma)\\
&-& (2\pi)^3\sum_{\Gamma'}\int\mbox{d}\mu(\Gamma')
\frac{\langle P_j|D^b(0)|\Gamma'\rangle\langle\Gamma' |D^{a}(0)|P_i\rangle}
{-\nu+E-E_{\Gamma'}-i\epsilon}\delta(\vec q-\vec P+\vec P_{\Gamma'})\nonumber\,.
\eea
Assuming that the target has isospin $I_3=+1/2$, the isospin-odd component
$F^{(-)}(\nu)$ can be extracted as $F^{(-)}(\nu) = \frac14(F^{1+i2,1-i2}(\nu)
-F^{1-i2,1+i2}(\nu))$.
Furthermore, noting from (\ref{8b}) that $F^{ab}(\nu)$ has no singularities 
in the upper half-plane $\nu$, the Cauchy theorem can be applied
on a closed contour extending along the real axis and closed in the upper
half-plane. This gives the dispersion relation\footnote{We used here the 
relation $F^{(-)}(-\nu) = -F^{(-)*}(\nu)$ which can be obtained from a 
simple examination of (\ref{8b}).}, assumed to require no subtraction
\bea\label{8c}
\mbox{Re }F^{(-)}(\nu) = \frac{2\nu}{\pi}{\cal P}\int_0^\infty
\mbox{d}\zeta \frac{\mbox{Im }F^{(-)}(\zeta)}{\zeta^2-\nu^2}\,.
\eea
By adding the imaginary part of $F^{(-)}(\nu)$ to both sides,
(\ref{8c}) can be rewritten as
\bea\label{8d}
F^{(-)}(\nu) = \frac{2\nu}{\pi}\int_0^\infty
\mbox{d}\zeta \frac{\mbox{Im }F^{(-)}(\zeta)}{\zeta^2-\nu^2}
\eea
where $\nu$ is understood to have a small positive imaginary part.

The imaginary part of $F^{(-)}(\nu)$ can be obtained from (\ref{8b}).
For $\nu>0$ only the first term will contribute (because the target is the
lowest-lying state) and the result can be expressed
 in terms of inclusive 
cross-sections for $\pi$ scattering on an antitriplet baryon as
\bea\label{10}
\mbox{Im }F^{(-)}(\nu) &=&
\frac18(2\pi)^4\sum_\Gamma\int\mbox{d}\mu(\Gamma)
|\langle P|D^{1+i2}|\Gamma\rangle|^2\delta(q+P-P_\Gamma)\\ 
&-&
\frac18(2\pi)^4\sum_{\Gamma'}\int\mbox{d}\mu(\Gamma')
|\langle P|D^{1-i2}|\Gamma'\rangle|^2\delta(q+P-P_{\Gamma'})\nonumber\\
&=&
\frac{f_\pi^2 \nu}{4}\left(\sigma_0(\pi^-\Xi^+_c\to X) -
\sigma_0 (\pi^+\Xi^+_c\to X)\right)\,.\nonumber
\eea
The cross-sections $\sigma_0$ correspond to off-shell incident
pions of momentum $q$ with $q^2=0$. This value of $q^2$ is needed
in order to be able to make use of the relation (\ref{8a}) on the
l.h.s. of the fixed-$q^2$ dispersion relation (\ref{8d}).

Inserting (\ref{8d}) and (\ref{10}) into (\ref{8a}) 
one obtains the well-known result for the 
Adler-Weisberger sum rule on an antitriplet baryon target
\bea\label{12}
1 = \frac{f_\pi^2}{\pi}\int_{m_\pi}^\infty \frac{\mbox{d}\nu}{\nu}
\left(\sigma_0(\pi^-\Xi^+_c\to X) -
\sigma_0(\pi^+\Xi^+_c\to X)\right)\,.
\eea

Let us assume in the following that the resonances dominate the integral in 
(\ref{12}), that is, the contribution of the continuum states can be 
neglected. We will estimate the error induced by this approximation later
in this section.
Then $\sigma_0(\pi^+\Xi^+_c\to X)$ vanishes as this state has 
isospin 3/2 and there are no heavy baryons with this quantum number.
Furthermore, the remaining cross-section in (\ref{12}) can be expressed in
terms of the pionic width (into off-shell pions with $q^2=0$) of the
respective excited state as
\bea\label{13}
\sigma_0(\pi^-\Xi^+_c\to X) = 2\pi^2\sum_{res}(2J+1)
\frac{\Gamma_0(X_{res}\to\pi^-\Xi^+_c)}{\nu^2}\delta(\nu-\delta 
M_{exc})\,.
\eea
Thus, the Adler-Weisberger sum rule on an ${\bf \bar 3}$ baryon target reads,
when only resonances are retained,
\bea\label{14}
1 &=& \pi f_\pi^2\sum_{res}(2J+1)\frac{\Gamma(X_{res}\to\pi^-\Xi^+_c)}
{\nu^3}\\
&=& \overbrace{\frac32 g_2^2}^{P-wave}
+ \overbrace{\frac12 h_3^2 + h_6^2 + \cdots}^{S-wave}
+ \overbrace{\frac43 h_{10}^2|\vec p_\pi\,|^2
+ \frac83 h_{14}^2|\vec p_\pi\,|^2
+ \cdots}^{D-wave}\,,\nonumber
\eea
where we have accounted explicitly for the contribution of the sextet 
s-wave baryons and of the p-wave baryons.
The ellipsis stand for
contributions from higher states which can decay to the
ground state ${\bf \bar 3}$ baryons with emission of one pion.
The pion momenta in (\ref{14}) are independent of the heavy quark
mass in the infinite mass limit, so the sum rule holds for any
species of heavy quark.

In a completely analogous way one can derive a sum rule involving the
coupling $g_1$ from the scattering amplitude of pions
off a sextet baryon. We define the corresponding retarded commutator
averaged over the baryon spin as in (\ref{8}) and take the baryon to be
a member of an isospin doublet with $I_3=1/2$ and spin 1/2. There is however,
a difference when compared with the previous case, due to the fact that now
there exist states lighter than the target: the s-wave ${\bf\bar 3}$ baryons.
As a result the two cuts of the function $F^{(-)}(\nu)$ along the real
axis touch each other and partly overlap. However, the dispersion relation
(\ref{8d}) remains valid, due to our deliberate choice of working
with the retarded commutator instead of the more usual time-ordered
product (see e.g. \cite{CL}).
In the case of the time-ordered product, the left-hand cut
(due to the second term in (\ref{8b})) sits above the real axis. This is no
problem as long as the two cuts do not touch, as the contour can be taken
to run above and below the cuts and close on a circle in the upper and lower
half-planes. When the cuts touch and overlap, such a choice of the contour
is not possible anymore. The method adopted here avoids these complications, as
the cuts are always under the real axis and they never get to pinch the contour.

Due to the presence of states lighter than the target, the second term
in (\ref{8b}) starts to contribute to Im $F^{(-)}(\nu)$ for positive $\nu$.
For this case, the relation (\ref{10}) is modified and reads (for $\nu>0$)
\bea\label{16}
\mbox{Im }F^{(-)}(\nu) &=& 
\frac18(2\pi)^4\sum_\Gamma\int\mbox{d}\mu(\Gamma)\left\{
|\langle P|D^{1+i2}|\Gamma_>\rangle|^2\delta(q+P-P_{\Gamma_>}) \right.\\
& &\qquad\left.+
|\langle P|D^{1+i2}|\Gamma_<\rangle|^2\delta(q-P+P_{\Gamma_<})\right\}\nonumber\\ 
&-&
\frac18(2\pi)^4\sum_{\Gamma'}\int\mbox{d}\mu(\Gamma')\left\{
|\langle P|D^{1-i2}|\Gamma'_>\rangle|^2\delta(q+P-P_{\Gamma'_>}) \right.\nonumber\\
& &\qquad\left.+
|\langle P|D^{1-i2}|\Gamma'_<\rangle|^2\delta(q-P+P_{\Gamma_<})\right\}\nonumber\,.
\eea
We denoted here by $\Gamma_>$ ($\Gamma_<$) the states lying above (below) the
target mass.
The sum over $\Gamma_>$
can be expressed as before in terms of inclusive cross-sections for $\pi$
scattering, and the one over $\Gamma_<$ can be computed in terms of the decay
width for the process ``target $\to \Gamma_<\pi$''
\bea\label{17}
\Gamma_0(T\to \Gamma_<\pi^+) = \frac{1}{2\pi}
\sum_{s_\Gamma}\nu |\langle\pi^+\Gamma_<|T\rangle|^2\,.
\eea
The contribution of the states $\Gamma$ which are degenerate with the
target will be extracted explicitly. There are two such states, the {\bf 6}
baryons with spins 1/2 and 3/2 (for the sum rule on a ${\bf\bar 3}$ baryon
this contribution vanished as pions do not couple to the ${\bf\bar 3}$
states). Their contributions on the left-hand side of (\ref{8a}) can be obtained
from (\ref{8b}) and are
\bea
\frac{\partial}{\partial\nu}F^{(-)}_{pole}(\nu)\vert_{\nu=0} = \frac{g_1^2}{8} +
\frac{g_1^2}{16} = \frac{3g_1^2}{16}\,.
\eea

The total contribution of all states which are not degenerate with the target to
(\ref{16}) can be written as
\bea\label{18}
\mbox{Im }F^{(-)}(\nu) &=& \frac12 f_\pi^2\pi^2\sum_{\Gamma_<}
\left\{\Gamma_0(T\to\pi^+\Gamma) - \Gamma_0(T\to\pi^-\Gamma)\right\}
\frac{1}{\nu} \delta(m_T-\nu-m_\Gamma)\\
& & + \frac{f_\pi^2\nu}{4}\left( \sigma_0(\pi^-T\to\Gamma) -
\sigma_0(\pi^+T\to\Gamma)\right)\,.\nonumber
\eea
Keeping, as before, just the one-body states as intermediate states,
one has (taking into account the fact that we have chosen the target $T$ to
have $I_3=+1/2$) that
$\Gamma(T\to\pi^-\Gamma)=0$ and $\sigma(\pi^+T\to \Gamma)=0$.
Inserting (\ref{18}) and (\ref{8d}) into (\ref{8a})
and keeping explicitly the contributions of the s-wave ${\bf\bar 3}$ and
p-wave baryons, one obtains the following form for the Adler-Weisberger sum
rule on a target {\bf 6} baryon
\bea\label{22}
1 &=& \overbrace{\frac{3g_1^2}{8} + \frac{g_2^2}{2}}^{P-wave}
+ \overbrace{\frac{h_2^2}{2} + \frac{h_4^2}{4}
+ \frac{h_5^2}{4} + \frac{h_7^2}{2} + \cdots}^{S-wave} \\
&+& \overbrace{\frac49 h_8^2|\vec p_\pi\,|^2 + \frac{2}{9}h_9^2 |\vec p_\pi\,|^2
+ \frac13 h_{11}^2|\vec p_\pi\,|^2 + \frac{2}{9}h_{12}^2|\vec p_\pi\,|^2 +
\frac49 h_{13}^2|\vec p_\pi\,|^2 + \frac23 h_{15}^2|\vec p_\pi\,|^2 +
\cdots}^{D-wave}\,.\nonumber
\eea
The ellipsis stand again for contributions from higher excited states
which can decay to the {\bf 6} baryons with emission of a single pion. 

Taking as target a polarized spin-3/2 sextet baryon gives new sum rules.
For spin projection $m_z=+1/2$ along the incident pion direction we obtain
\bea\label{+1/2}
1 &=& \overbrace{\frac{3g_1^2}{16} + g_2^2}^{P-wave} + 
\overbrace{\frac{h_2^2}{2} + \frac{h_4^2}{4}
+ \frac{h_5^2}{4} + \frac{h_7^2}{2} + \cdots}^{S-wave} \\
&+& \overbrace{\frac23 h_8^2|\vec p_\pi\,|^2 + \frac13 h_9^2 |\vec p_\pi\,|^2
+ \frac16 h_{11}^2|\vec p_\pi\,|^2 + \frac13 h_{12}^2 |\vec p_\pi\,|^2
+ \frac23 h_{13}^2 |\vec p_\pi\,|^2 + \frac13 h_{15}^2 |\vec p_\pi\,|^2 +
\cdots}^{D-wave}\,,\nonumber
\eea
and for $m_z=+3/2$
\bea\label{+3/2}
1 &=& \overbrace{\frac{9g_1^2}{16}}^{P-wave} +
\overbrace{\frac{h_2^2}{2} + \frac{h_4^2}{4}
+ \frac{h_5^2}{4} + \frac{h_7^2}{2} + \cdots}^{S-wave} \\
&+& \overbrace{\frac29 h_8^2|\vec p_\pi\,|^2 + \frac19 h_9^2 |\vec p_\pi\,|^2
+ \frac12 h_{11}^2|\vec p_\pi\,|^2 + \frac19 h_{12}^2|\vec p_\pi\,|^2
+ \frac29 h_{13}^2 |\vec p_\pi\,|^2 + h_{15}^2|\vec p_\pi\,|^2 +
\cdots}^{D-wave}\,,\nonumber
\eea
In fact only one of these sum rules is new: by taking their average the
unpolarized sum rule (\ref{22}) is recovered. We will take as the new
independent
sum rule the difference of (\ref{+1/2}) and (\ref{+3/2}) written as
\bea\label{1/2-3/2}
\frac{3g_1^2}{8} - g_2^2 &=&
\frac49 h_8^2|\vec p_\pi\,|^2 + \frac29 h_9^2 |\vec p_\pi\,|^2
- \frac13 h_{11}^2|\vec p_\pi\,|^2 + \frac29 h_{12}^2|\vec p_\pi\,|^2\\
& &\qquad
+ \frac49 h_{13}^2 |\vec p_\pi\,|^2 - \frac23 h_{15}^2|\vec p_\pi\,|^2 +
\cdots\,.\nonumber
\eea
One can see that the contributions of the S-wave couplings have canceled
out in taking the difference. The phenomenological consequences of
this sum rule will be discussed in Section VI.

\subsection{Continuum contributions}

We have neglected in the above considerations the contributions to the
sum rule from continuum states. This is likely to be a good approximation
in nature, where the heavy baryons are seen as narrow states
with widths much smaller than their mass separation. A similar approximation
has been justified in the meson case \cite{CP} by using large-$N_c$
arguments: the contribution of 2-body states to the sum rule is
suppressed relative to the one of the resonances by $1/N_c$.
The situation in the baryon case is, however, completely different.
In the following we will enumerate the contributions of a few
intermediate states to the sum rule in the large-$N_c$ limit,
following \cite{Wi,Co}.

The coupling of an s-wave heavy baryon to the Goldstone bosons
scales as $\sqrt{N_c}$, which can be understood by recalling that the
pion can couple to each of the $N_c-1$ light quarks in the
baryon (the factor of $1/f_\pi$ gives an additional suppression of
$1/\sqrt{N_c}$). This gives that $g_1$ and $g_2$ scale as $N_c$.
The similar s-wave to p-wave couplings are however only of order 1,
because the Goldstone boson can only couple to the quark in a p-wave
state. This implies that the couplings of the p-wave states $h_i$ scale
like $\sqrt{N_c}$.

On the other hand, the amplitude for the process pion +
s-wave baryon $\to$ pion + s-wave baryon is also of order 1 \cite{Wi,Co}.
This means that the 2-body states ($\pi$, s-wave baryon) contribute
to the sum rule at the same order in $1/N_c$ as the 1-body states
with p-wave baryons. This fact could potentially upset the resonance
saturation approximation of the sum rules made above.
Therefore an estimate of the continuum contribution is necessary. 

We will restrict ourselves to the study of the continuum
contributions to the sum rule on a {\bf 3} baryon (\ref{14}).
Even without an explicit calculation it can be argued, as
in the heavy meson case \cite{CP}, that the continuum 
contribution must be positive since there are
more states containing one heavy quark with isospin 1/2 than 3/2.
Our explicit calculations will confirm this conjecture,
at least for the low-energy region where we can compute
the continuum contribution. This implies that the sum rules 
(\ref{14}) and (\ref{22}-\ref{1/2-3/2}) should in fact be 
considered as inequalities.

In the absence of experimental data, the only reliable
information we have about the continuum contribution comes
from chiral perturbation theory ($\chi PT$). Unfortunately its 
validity is restricted to the low-energy region in the vicinity of 
the threshold for $\pi-{\bf\bar 3}$ scattering. 
In this subsection we compute the continuum contribution
from threshold up to the cut-off $\Lambda=345$ MeV, which will
be shown to mark the limit of validity of $\chi$PT in this
system. In the next section (IV.C.) we go beyond $\chi$PT 
and use unitarity to bound the total contributions
of the S-wave and P-wave channels to the AW sum rule.

The contribution of the ($\pi$, s-wave baryon) continuum to the AW
sum rule on a {\bf 3} baryon (\ref{14}) is expressed in terms of
the cross-sections appearing in (\ref{12}). In accordance with our
previous
discussion we keep only the contributions of the following channels
\bea
\sigma_-(\nu) &=& \sigma_0(\pi^-\Xi_c^+\to (\pi {\bf \bar 3})_{1/2}) +
\sigma_0(\pi^-\Xi_c^+\to (\pi {\bf \bar 3})_{3/2}) \\
& &\,+ \sigma_0(\pi^-\Xi_c^+\to (\pi {\bf 6})_{1/2}) +
\sigma_0(\pi^-\Xi_c^+\to (\pi {\bf 6})_{3/2})\nonumber\\
\sigma_+(\nu) &=& \sigma_0(\pi^+\Xi_c^+\to (\pi {\bf \bar 3})_{3/2}) +
\sigma_0(\pi^+\Xi_c^+\to (\pi {\bf 6})_{3/2})\,.
\eea
We have separated in $\sigma_-(\nu)$ the contributions of the continuum
states with isospins I=1/2 and 3/2. The corresponding amplitudes are given
by the usual rules for isospin addition as
\bea
{\cal M}(\pi^-\Xi_c^+\to (\pi {\bf \bar 3})_{1/2}) &=&
\frac{1}{\sqrt3}{\cal M}(\pi^-\Xi_c^+\to \pi^0\Xi_c^0 )  -
\sqrt{\frac23} {\cal M}(\pi^-\Xi_c^+\to \pi^-\Xi_c^+ )\\
{\cal M}(\pi^-\Xi_c^+\to (\pi {\bf \bar 3})_{3/2}) &=&
\sqrt{\frac23} {\cal M}(\pi^-\Xi_c^+\to \pi^0\Xi_c^0 )  +
\frac{1}{\sqrt3} {\cal M}(\pi^-\Xi_c^+\to \pi^-\Xi_c^+ )\,.
\eea
The evaluation of the diagrams shown in Fig.1 gives
\bea\label{ampl1}
{\cal M}(\pi^-\Xi_c^+\to (\pi {\bf \bar 3})_{1/2}) &=&
-\frac{i}{2}\sqrt{\frac32}\frac{1}{f_\pi^2}\bar u(v,s')\left\{
\frac83 v\cdot q\right.\\
& &\hspace{-2cm}\left.\, + g_2^2 [(v\cdot q)^2-p\cdot q]
\left(\frac{1}{-v\cdot q-\Delta+i\Gamma_{6^*}/2} -
\frac{3}{v\cdot q-\Delta+i\Gamma_{6^*}/2}\right)\right\}u(v,s)
\nonumber\\
{\cal M}(\pi^-\Xi_c^+\to (\pi {\bf \bar 3})_{3/2}) &=&
-\frac{i\sqrt3}{2f_\pi^2}\bar u(v,s')\left\{
\frac23 v\cdot q +
g_2^2\frac{(v\cdot q)^2-p\cdot q}{-v\cdot q-\Delta+i\Gamma_{6^*}/2}
\right\}u(v,s)\\\label{ampl3}
{\cal M}(\pi^+\Xi_c^+\to (\pi {\bf \bar 3})_{3/2}) &=&
-\frac{i}{2f_\pi^2}\bar u(v,s')\left\{
2v\cdot q + 3g_2^2 \frac{(v\cdot q)^2-p\cdot q}{-v\cdot q-\Delta+i\Gamma_6/2}
\right\}u(v,s)\,.
\eea

Note that, as explained above, the incoming pion has $q^2=0$; however,
the final one is on the mass-shell $p^2=m_\pi^2$.
We have denoted here $\Delta=M_{\bf 6}-M_{\bf \bar 3}$, the mass splitting
between the ${\bf \bar 3}$ and {\bf 6} multiplets. The widths in the
denominators include both the charged and neutral pion channels and
correspond to on-shell final pions.  In the
heavy mass limit they are equal and are given by
\bea
\Gamma_6 = \Gamma_{6^*} = \frac{3g_2^2}{8\pi f_\pi^2}(\Delta^2-m_\pi^2)^{3/2}\,.
\eea

The amplitudes (\ref{ampl1}-\ref{ampl3}) give, after squaring and 
integrating over the phase space of the final pion, the following 
cross-sections
\bea\label{xsec1}
\sigma_0(\pi^-\Xi_c^+\to (\pi {\bf \bar 3})_{1/2}) &=&
\frac{3}{64\pi f_\pi^4}\left\{ \frac{128}{9}E_\pi |\vec p_\pi\,|\right.\\
& &\hspace{-2cm}\left. + \,
\frac23 g_2^4 E_\pi |\vec p_\pi\,|^3 \left|
\frac{1}{-E_\pi-\Delta+i\Gamma_{6^*}/2} -
\frac{3}{E_\pi-\Delta+i\Gamma_{6^*}/2} \right|^2\right\}\nonumber\\
\sigma_0(\pi^-\Xi_c^+\to (\pi {\bf \bar 3})_{3/2}) &=&
\frac{3}{32\pi f_\pi^4}\left\{ \frac89 E_\pi|\vec p_\pi\,| + \frac23 g_2^4 
E_\pi |\vec p_\pi\,|^3 \frac{1}{(E_\pi+\Delta)^2+\Gamma_{6^*}^2/4}\right\}\\
\label{xsec3}
\sigma_0(\pi^+\Xi_c^+\to (\pi {\bf \bar 3})_{3/2}) &=&
\frac{1}{32\pi f_\pi^4}\left\{ 8E_\pi |\vec p_\pi\,| + 6g_2^4 
E_\pi |\vec p_\pi\,|^3 \frac{1}{(E_\pi+\Delta)^2+\Gamma_6^2/4}\right\}\,.
\eea

An important point which must be taken into account is that the
expression (\ref{xsec1}) contains a resonant piece. Upon insertion
in the AW sum rule (\ref{12}), it reproduces, in the narrow width
approximation, the contribution of the 1-body states with a sextet
baryon shown in (\ref{14}). Therefore, to avoid double counting of this
term, the true continuum contribution to the sum rule is obtained by
explicitly subtracting it. To see this explicitly, we insert the
resonant term in (\ref{xsec1}) into the integral on the r.h.s. of the
AW sum rule. We obtain in the narrow width approximation
\bea
I_{AW} &=& \frac{f_\pi^2}{\pi}\int\frac{\mbox{d}\nu}{\nu}
\sigma_-^{res}(\nu) =
\frac{9g_2^4}{32\pi^2 f_\pi^2}\int \mbox{d}\nu
\frac{(\nu^2-m_\pi^2)^{3/2}}{(\nu-\Delta)^2+\Gamma_6^2/4}
\to \frac32 g_2^2
\eea
as $\Gamma\to 0$.
In the last step we used the well-known representation of the $\delta$
function
\bea
\frac{\Gamma}{2\pi}\frac{1}{(\nu-\Delta)^2+\Gamma^2/4} \to
\delta(\nu-\Delta)\,,\qquad (\Gamma\to 0)\,.
\eea

The remaining cross-sections needed for the continuum corrections
to the sum rule (\ref{14}) correspond to final states with a sextet
baryon of spin 1/2 and 3/2 plus one pion. They can be computed
analogously with the result 
\bea\label{crs1}
\sigma_0(\pi^-\Xi_c^+\to \pi {\bf 6}) &=& 
\frac{g_1^2g_2^2}{128\pi f_\pi^4} E_\pi (E_\pi^{'2}-m_\pi^2)^{3/2}
\left|\frac{1}{-E'_\pi-\Delta+i\Gamma_6/2} + 
\frac{3}{E_\pi-\Delta+i\Gamma_6/2}
\right|^2\nonumber\\
& & +\, \frac{g_1^2g_2^2}{64\pi f_\pi^4}
\frac{E_\pi (E_\pi^{'2}-m_\pi^2)^{3/2}}{(E'_\pi+\Delta)^2+\Gamma_6^2/4}\,,
\qquad E'_\pi=E_\pi-\Delta\\
\sigma_0(\pi^+\Xi_c^+\to \pi {\bf 6}) &=& 
\frac{3g_1^2g_2^2}{64\pi f_\pi^4}
\frac{E_\pi (E_\pi^{'2}-m_\pi^2)^{3/2}}{(E'_\pi+\Delta)^2+\Gamma_6^2/4}\\
\sigma_0(\pi^-\Xi_c^+\to \pi {\bf 6}^*) &=& 
2\sigma_0(\pi^-\Xi_c^+\to \pi {\bf 6})\\
\sigma_0(\pi^+\Xi_c^+\to \pi {\bf 6}^*) &=&
2\sigma_0(\pi^+\Xi_c^+\to \pi {\bf 6})\,.
\eea
The first (second) term in (\ref{crs1}) corresponds to final states with
isospin 1/2 (3/2).

These cross-sections are inserted in the AW sum rule (\ref{12}) and 
the integration is done numerically. We included also the
first term in the elastic cross-section (\ref{xsec1}) although it is formally
of higher order in $1/N_c$ than the contributions we are interested in.
In fact it will be seen to dominate the continuum contribution.
The mass splitting between the ${\bf\bar 3}$ and {\bf 6} multiplets
will be taken $\Delta=225$ MeV, corresponding to the charmed baryons'
case. The quark model values (\ref{cqm}) for the couplings $g_1$ and $g_2$
will be used (with $g_A=0.75$). 
We obtain for the continuum contribution to the AW sum rule (\ref{14})
for a few values of the upper limit of integration the numbers shown in
Table 2. Especially for larger values of the upper cut-off, these
corrections appear as being significant when compared with the 1-body
sextet contribution to the sum rule $3/2 g_2^2=0.562$.

\begin{center}
\begin{tabular}{|c|ccccccc|}
\hline
& \multicolumn{7}{|c|}{Cut-off energy (MeV)}\\
\cline{2-8}
 & 345 & 370 & 395 & 420 & 445 & 470 & 495 \\
\hline
\hline
$({\bf\bar 3}+\pi)_S$ & 0.117 & 0.141 & 0.167 & 0.195 & 0.225 & 0.257 & 0.290 \\
\hline
$({\bf\bar 3}+\pi)_P$ & 0.020 & 0.033 & 0.045 & 0.057 & 0.068 & 0.078 & 0.089 \\
\hline
$({\bf 6}+\pi)_P$ & 0.000 & 0.000 & 0.000 & 0.001 & 0.001 & 0.002 & 0.003 \\
\hline
Total & 0.137 & 0.174 & 0.213 & 0.253 & 0.294 & 0.337 & 0.382 \\
\hline
\end{tabular}
\end{center}
\begin{quote} {\bf Table 2.}
Continuum contributions to the AW sum rule on a ${\bf\bar 3}$ baryon
corresponding to different partial waves.
\end{quote}

In reality we will see that these large contributions are simply an effect
of the limited applicability of chiral perturbation theory for the
particular process of pion scattering on a static heavy baryon.
It will be shown in the following section that unitarity gives an
upper bound on the elastic pion scattering cross sections, which is
exceeded in the $L=0$, $I=1/2$ channel already above $E_\pi=345$ MeV.
Therefore the only trustworthy values in Table 2 are those corresponding
to the lowest value of the cut-off.

\subsection{Unitarity constraints}

The amplitude for elastic pion scattering $\pi^a T_i\to \pi^b T_j$
on a static target $T$ with
isospin 1/2  can be written as in (\ref{8}) in terms of
two functions $T^{(\pm)}$ 
\bea
T^{ba}_{ji} = T^{(+)}\delta_{ij}\delta_{ab} + 
T^{(-)}\frac12 [\tau^a,\,\tau^b]_{ji}\,.
\eea
Each of these functions can be expressed in terms of two amplitudes $f^{(\pm)}$
and $g^{(\pm)}$, corresponding to spin-nonflip and respectively spin-flip 
transitions, defined by
\bea
T^{(\pm)}(E_\pi,\,\cos\theta) = u^\dagger(v,s')\left( f^{(\pm)}(E_\pi,\,\cos\theta)
+ ig^{(\pm)}(E_\pi,\,\cos\theta)\vec\sigma\cdot(\hat q'\times \hat q)\right)
u(v,s)\,.
\eea
We denoted here by $E_\pi$ and $\cos\theta$ the pion energy and scattering
angle in the rest frame of the target. $\hat q,\,\hat q'$ are unit vectors
along the directions of the initial and final pions and $u(v,s)$ are 
nonrelativistic 2-spinors in the rest frame of T.

 The functions $f^{(\pm)}$ and $g^{(\pm)}$ have partial wave expansions of
the form (see e.g. \cite{CGLN})
\bea\label{pwef}
f^{(\pm)}(E_\pi,\,\cos\theta) &=& \sum_{l=0}^\infty
[(l+1)f_{l+}^{(\pm)}(E_\pi)+lf_{l-}^{(\pm)}(E_\pi)]P_l(\cos\theta)\\\label{pweg}
g^{(\pm)}(E_\pi,\,\cos\theta) &=& \sum_{l=1}^\infty 
[f_{l-}^{(\pm)}(E_\pi)-f_{l+}^{(\pm)}(E_\pi)]P'_l(\cos\theta)\,.
\eea

 From a physical point of view it is more transparent to work instead
of the amplitudes $T^{(\pm)}$ with amplitudes of well-defined isospin,
given by
\bea
T^{(1/2)} &=& T^{(+)} - 2T^{(-)}\\
T^{(3/2)} &=& T^{(+)} + T^{(-)}\,.
\eea
These amplitudes have partial-wave expansions similar to those in
(\ref{pwef}-\ref{pweg}). The corresponding amplitudes
will be called $f_{l\pm}^{(I)}(E_\pi)$, with $I=1/2,3/2$ and have
physical interpretation
of scattering amplitudes in channels with total angular momentum
$j=l\pm\frac12$, parity $-(-1)^l$ and isospin $I$.


Unitarity imposes an well-known constraint on the partial wave amplitudes
\bea\label{unitarity}
\mbox{Im }f_{l\pm}^{(I)} \geq \frac{|\vec p_\pi\,|}{4\pi}|f_{l\pm}^{(I)}|^2\,.
\eea
For small energies where only elastic scattering is allowed, the
inequality turns into an equality. The partial wave amplitudes are
usually parametrized in this region in terms of phase shifts
$\delta_{l\pm}^{(I)}$ as
\bea
f_{l\pm}^{(I)} = \frac{4\pi}{|\vec p_\pi\,|}\sin\delta_{l\pm}^{(I)}
e^{i\delta_{l\pm}^{(I)}}\,.
\eea

  From (\ref{unitarity}) one can derive an upper bound on the partial
wave amplitudes, valid both in the elastic and inelastic cases,
\bea\label{unitbound}
|f_{l\pm}^{(I)}| \leq \frac{4\pi}{|\vec p_\pi\,|}\,.
\eea

 We will use unitarity to derive an upper bound on the 
inclusive cross-sections into all possible final states.
The method makes use of the optical theorem which expresses
an inclusive cross-section in terms of the imaginary part
of a forward scattering amplitude.
Explicitly this gives the following expressions for
the inclusive cross-sections for pions incident on the
$I_3=+1/2$ member of an isospin doublet
\bea\label{x-sec1}
\sigma(\pi^- T\to X) &=& \frac{1}{|\vec p_\pi|}\mbox{Im }
{\cal M}(\pi^- T\to\pi^- T) = \frac{1}{3|\vec p_\pi|}\mbox{Im }
(T^{(3/2)}(E_\pi,1) + 2T^{(1/2)}(E_\pi,1) )\\\label{x-sec2}
\sigma(\pi^+ T\to X) &=& \frac{1}{|\vec p_\pi|}\mbox{Im }
{\cal M}(\pi^+ T\to\pi^+ T) = \frac{1}{|\vec p_\pi|}\mbox{Im }
T^{(3/2)}(E_\pi,1)\,.
\eea
Inserting here the partial wave expansions (\ref{pwef}-\ref{pweg})
gives
\bea
\sigma(\pi^- T\to X) &=& \frac{1}{3|\vec p_\pi|}
\sum_{l=0}^\infty [(l+1)\mbox{Im }(f_{l+}^{(3/2)}+2f_{l+}^{(1/2)}) +
l\mbox{Im }(f_{l-}^{(3/2)}+2f_{l-}^{(1/2)})]\\
\sigma(\pi^+ T\to X) &=& \frac{1}{|\vec p_\pi|}
\sum_{l=0}^\infty [(l+1)\mbox{Im }f_{l+}^{(3/2)} +
l\mbox{Im }f_{l-}^{(3/2)}]\,.
\eea
Taking the difference  we obtain
\bea\label{x-secdiff}
\sigma(\pi^- T\to X) - \sigma(\pi^+ T\to X) &=&\\
& &\hspace{-2cm}
\frac{2}{3|\vec p_\pi|}\sum_{l=0}^\infty
[(l+1)\mbox{Im }(f_{l+}^{(1/2)}-f_{l+}^{(3/2)}) +
l\mbox{Im }(f_{l-}^{(1/2)}-f_{l-}^{(3/2)})]\,.\nonumber
\eea

  From (\ref{unitarity}) one can see that Im $f_{l\pm}^{(I)}$ 
is positive and bounded from above by (\ref{unitbound}).
Therefore our strategy in the following will be to set
an absolute upper bound on the difference (\ref{x-secdiff}) by using
(\ref{unitbound}) for Im $f_{l\pm}^{(1/2)}$ and taking
Im $f_{l\pm}^{(3/2)}=0$ everywhere outside the domain of
applicability of chiral perturbation theory.

Strictly speaking the cross-sections appearing in the AW sum
rule are not the physical on-shell cross-sections 
(\ref{x-sec1},\ref{x-sec2}) but rather, cross-sections
with a massless incident pion. Let us explicitly write the dependence
of the partial wave amplitudes 
$f_{l\pm}^{(I)}(E_\pi;m_{\pi f},m_{\pi i})$ on the masses of the 
initial and final pions. Then the correct amplitudes to be used
in (\ref{x-secdiff}) would be $f_{l\pm}^{(I)}(E_\pi;0,0)$.
Unfortunately, no unitarity bound can be written for the
absolute value of this amplitude. To see this, we write the
inequality (\ref{unitarity}) keeping explicit the pion mass dependence
\bea\label{unoff-shell}
\mbox{Im }f_{l\pm}^{(I)}(E_\pi;0,0) \geq 
\frac{|\vec p_\pi\,|}{4\pi}|f_{l\pm}^{(I)}(E_\pi;m_\pi,0)|^2\,.
\eea
The quantities on the two sides of this inequality are different
and thus no useful information can be extracted about them.
We will use nevertheless the physical on-shell partial waves
in (\ref{x-secdiff}), as the pion mass effects can be expected
to be less important in the high-energy region where this
relation will be applied. 

At low energies we will use the lowest order chiral perturbation
theory result
obtained from an evaluation of the graphs in Fig.1. Performing a
partial wave decomposition we find that only the $l=0,1$ waves
are present at this order. Furthermore, there is no
spin-flip, which gives $f_{l+} = f_{l-} = f_l$. We obtain
\bea\label{pw1}
f_0^{(1/2)}(E_\pi) &=& \frac{2}{f_\pi^2}E_\pi\\\label{pw2}
f_1^{(1/2)}(E_\pi) &=& \frac{g_2^2}{4f_\pi^2}|\vec p_\pi\,|^2
\left(\frac{1}{-E_\pi-\Delta+i\Gamma_6/2} - 
\frac{3}{E_\pi-\Delta+i\Gamma_6/2}\right)\\\label{pw3}
f_0^{(3/2)}(E_\pi) &=& -\frac{1}{f_\pi^2}E_\pi\\\label{pw4}
f_1^{(3/2)}(E_\pi) &=& -\frac{g_2^2}{2f_\pi^2}|\vec p_\pi\,|^2
\frac{1}{-E_\pi-\Delta+i\Gamma_6/2}\,.
\eea

   From these expressions one can compute the pion 
energy $(E_\pi)_{max}=\Lambda_l^{(I)}$
at which the unitarity bound (\ref{unitbound}) is reached in each channel.
We obtain $\Lambda_S^{(1/2)}=345$ MeV, $\Lambda_P^{(1/2)}=449$ MeV,
$\Lambda_S^{(3/2)}=477.8$ MeV and $\Lambda_P^{(3/2)}=1189.5$ MeV.
We used in (\ref{pw2},\ref{pw4}) the same values for $\Delta$
and $g_2$ as in our previous estimates. We will consider in the
following these values as marking the limits of validity of chiral
perturbation theory in each channel.

We can compute now the contributions to the right-hand side of
the AW sum rule for each channel, following the prescription
outlined above.

a) $I=1/2$, $S$-wave. Below the unitarity limit $\Lambda_S^{(1/2)}$
we insert the $\chi PT$ results for the cross-sections
(\ref{xsec1},\ref{crs1}) into the AW sum rule. We obtain
\bea
I_{S_1}^{(1/2)} = \frac{f_\pi^2}{\pi}\int_{m_\pi}^{\Lambda_S^{(1/2)}}
\frac{\mbox{d}\nu}{\nu}\sigma_0(\pi^- T\to (\pi T)_{1/2})_S
 = 0.156\,.
\eea
Above $\Lambda_S^{(1/2)}$ we use the unitarity limit (\ref{unitbound})
\bea
I_{S_2}^{(1/2)} \leq \frac{f_\pi^2}{\pi}\int_{\Lambda_S^{(1/2)}}^\infty
\frac{\mbox{d}\nu}{\nu}
\frac{8\pi}{3(\nu^2-m_\pi^2)} = 0.212\,.
\eea
Adding these two contribution we obtain an upper limit on the contribution
of this channel to the AW sum rule
\bea\label{Swave1/2}
I_S^{(1/2)} = I_{S_1}^{(1/2)} + I_{S_2}^{(1/2)} \leq 0.368\,.
\eea

b) $I=1/2$, $P$-wave. The contribution of this channel to the AW
sum rule can be split up as in the previous case into two terms.
The low-energy part contains contributions from the both possible
final states $(\pi {\bf\bar 3})$ and $(\pi {\bf 6})$
\bea\label{IP1low}
I_{P_1}^{(1/2)} &=& \frac{f_\pi^2}{\pi}\int_{m_\pi}^{\Lambda_P^{(1/2)}}
\frac{\mbox{d}\nu}{\nu}\left(\sigma_0(\pi^- T\to (\pi T)_{1/2})_P +
\sigma_0(\pi^- T\to (\pi S)_{1/2})_P\right)\\
&=& \frac32g_2^2 + 0.072 + 0.001 = \frac32g_2^2 + 0.073\,.\nonumber
\eea
The contribution of the 1-body state with a sextet s-wave baryon has
been extracted explicitly, as discussed above (see the paragraph following
Eq.(\ref{xsec3})).
Above $E_\pi=\Lambda_P^{(1/2)}$ we use again the unitarity limit
\bea
I_{P_2}^{(1/2)} \leq \frac{f_\pi^2}{\pi}\int_{\Lambda_P^{(1/2)}}^\infty
\frac{\mbox{d}\nu}{\nu}\,3
\frac{8\pi}{3(\nu^2-m_\pi^2)} = 0.362\,.
\eea
Taking the sum gives
\bea\label{Pwave1/2}
I_P^{(1/2)} = I_{P_1}^{(1/2)} + I_{P_2}^{(1/2)} \leq
\frac32g_2^2 + 0.435\,.
\eea

c) $I=3/2$, $S$-wave. For the $I=3/2$ channel we include,
as discussed above, only the contribution of the low-energy region,
so as to maximize the integral appearing in the AW
sum rule. The $S$-wave contributes
\bea
I_S^{(3/2)} = -\frac{f_\pi^2}{\pi}\int_{m_\pi}^{\Lambda_S^{(3/2)}}
\frac{\mbox{d}\nu}{\nu}\left(\sigma_0(\pi^- T\to (\pi T)_{3/2})_S
- \sigma_0(\pi^+ T\to (\pi T)_{3/2})_S\right)
 = -0.089\,.
\eea

d) $I=3/2$, $P$-wave. This channel receives significant contributions
only from the $(\pi {\bf\bar 3})$ final states
\bea
I_P^{(3/2)} = -\frac{f_\pi^2}{\pi}\int_{m_\pi}^{\Lambda_P^{(3/2)}}
\frac{\mbox{d}\nu}{\nu}\left(\sigma_0(\pi^- T\to (\pi T)_{3/2})_P
- \sigma_0(\pi^+ T\to (\pi T)_{3/2})_P\right)
= -0.039\,.
\eea
The states $(\pi {\bf 6})$ give a very small contribution, of
$-0.0004$.

The quantity on the l.h.s. of the bound (\ref{Swave1/2}) includes,
just as in (\ref{IP1low}), also contributions
from 1-body states. These states are heavy baryons which can decay
to the s-wave ${\bf \bar 3}$ baryon by $S$-wave pion emission.
Their contribution to the sum rule has been given in (\ref{14}).
Combining (\ref{14}) with the limit (\ref{Swave1/2}) gives the
constraint
\bea\label{SBound}
\frac12 h_3^2 + h_6^2 + (\mbox{S-wave, I=1/2 continuum}) \leq 0.368\,.
\eea

Taken alone, this inequality tells us absolutely nothing about the
couplings $h_3$ and $h_6$. To do so, it must be supplemented
with additional information about the continuum.
For example, assuming that the continuum contribution is positive,
(\ref{SBound}) gives an upper bound on these 
couplings\footnote{A similar reasoning applied to pion-heavy meson
scattering gives the inequality $h^2$ + (S-wave, I=1/2 continuum)
$\leq 0.368$ (with the notations of \cite{CP}). In this case however
we can invoke large-$N_c$ arguments to argue that the continuum
is suppressed.}.
Experience with the lowest order $\chi PT$ results shows
that this is very likely the case.

A similar prediction can be made about the P-wave $I=1/2$ channel.
The relation (\ref{Pwave1/2}) limits the contribution of this channel
to the sum rule.
\bea\label{PBound}
I_P^{(1/2)}  \leq  \frac32g_2^2 + 0.396\,.
\eea
The only bound states (besides the s-wave {\bf 6} whose contribution
is made explicit) contributing in this channel are radial
excitations of s-wave baryons. The expression (\ref{PBound})
gives an upper bound on their contribution plus continuum
to the sum rule.
Note, however, that in contrast to the bound (\ref{SBound})
which is parameter-free, the numerical bound in (\ref{PBound}) has
been obtained with the quark model value for $g_2$.

Unfortunately, not much can be said with the help of these
methods about the $D$-wave contributions to the sum rule.
We do not have any control over the low energy
behaviour of these  partial waves, which would require a
next-to-leading calculation in chiral perturbation theory.
Also, the use of unitarity for the high-energy region gives
an upper bound which is too large to be useful. 


\section{Constituent quark model predictions for strong decays of heavy baryons}

So far everything has been completely general and the sum rules
(\ref{14}) and (\ref{22}-\ref{1/2-3/2}) are model-independent. We 
specialize now to
the case of the constituent quark model, where some of the couplings
are related. For example, the couplings of the s-wave baryons are given in
the quark model by \cite{yan}
\bea\label{cqm}
g_1 = -\frac43 g_A\,,\qquad |g_2| = \sqrt{\frac23}g_A
\eea
with $g_A\simeq 0.75$ the constituent pion-quark coupling.
The strong and electromagnetic decays of the charmed baryons
have been also studied in the quark model with SU(6) symmetry
in \cite{DL}.

We will show in the following that the S-wave couplings of the p-wave 
baryons are also related in the quark model as
\bea\label{23}
\frac{|h_3|}{|h_4|} = \frac{\sqrt{3}}{2}\,,\qquad 
\frac{|h_2|}{|h_4|} = \frac{1}{2}\,,\qquad
h_5 = h_6 = h_7 = 0\,.
\eea

One possible way of deriving these relations is by comparing the matrix
elements of the axial current between s-wave and p-wave baryons computed
in two different ways. First, the axial vector current in the effective
theory is given by the Noether theorem. To the lowest order in the pion
field, it is given by the coefficient of $-(\sqrt2/f_\pi)\partial_\mu
\pi^a$ in the interaction Lagrangian. For (\ref{6}) this gives
\bea
J_\mu^a &=& h_2\epsilon_{ijk} \bar S^{kl}_\nu v_\mu t^a_{lj}R^i_\nu +
h_3\mbox{tr }(\bar B_{\bar 3}v_\mu t^a U) +
h_4\mbox{tr }(\bar S_\nu v_\mu t^a V_\nu )\\
&+& h_5\mbox{tr }(\bar S_\nu v_\mu t^a R'_\nu ) +
h_6 \bar T_i v_\mu t^a_{ji}U'_j +
h_7\epsilon_{ijk} \bar S^\nu_{kl}v_\mu t^a_{lj}V^{'i}_\nu +
\mbox{pion terms}\nonumber\,.
\eea
We denoted here $t^a=\lambda^a/2$, the generators of the diagonal flavor 
SU(3) group. One can see that only the matrix elements of the $\mu=0$
component are nonvanishing in the hadron rest frame.

On the other hand, the matrix elements of $J^a_0$ can be computed between
the corresponding constituent quark model states. In the nonrelativistic
limit, the matrix element of $J^{1-i2}_0$ can be written as
\bea\label{AtimeNR}
\langle f|\bar d\gamma_0\gamma_5 u|i\rangle =
\langle f|\frac{\vec\sigma\cdot\vec p}{2m_u}+
          \frac{\vec\sigma\cdot\vec p\,'}{2m_d}|i\rangle\,,
\eea
where $m_u,m_d$ are the constituent quark masses, assumed for simplicity equal
in the following. The $\sigma$ matrices act only on the spin of the quark
changing its flavor $u\to d$.

The relations in (\ref{23}) are obtained from considering the following
matrix elements in the limit $q\to 0$
\bea\label{24}
& &\langle\Sigma_{c}^0\uparrow | \bar d\gamma_0
\gamma_5 u|\Lambda_{c1}^{+}(\frac12)\uparrow\rangle = -h_2\\
& &\langle\Lambda_{c}^+\uparrow | \bar d\gamma_0
\gamma_5 u|\Sigma_{c0}^{++}(\frac12)\uparrow\rangle = h_3\\
\label{25}
& &\langle\Sigma_{c}^+\uparrow | \bar d\gamma_0
\gamma_5 u|\Sigma_{c1}^{++}(\frac12)\uparrow\rangle = -\frac{1}{\sqrt2}h_4\,.
\eea

One expects these matrix elements to be related in the quark model,
as the baryon states on the l.h.s. have the same orbital 
wavefunction (as do the baryons on the r.h.s.). The flavor-spin
wavefunctions of the baryon states shown are given in the Appendix.
The p-wave states are represented as linear combinations of basis vectors
$|c(m_Q)L(m_\ell)q(m_1)q(m_2)\rangle$ with $m_Q$ the $z$-projection of the
heavy quark spin, $m_\ell$ the projection of the total orbital angular
momentum and $m_{1,2}$ the projections of the light quarks' spins.
The operators in (\ref{AtimeNR}) can be written in terms of spin-raising,
spin-lowering operators and $\sigma_3$ as
\bea
\vec\sigma\cdot\vec p = \sqrt2\sigma_+ p_- - \sqrt2\sigma_- p_+ + \sigma_3 p_z
\eea
where $(p_-,p_z,p_+)$ form the components of a $J=1$ spherical tensor.
The matrix element (\ref{AtimeNR}) can be then simply evaluated with the help
of the Wigner-Eckart theorem. Using the wavefunctions (\ref{A1}-\ref{A4}) one
obtains
\bea\label{30}
& &\langle\Sigma_{c}^0\uparrow |\vec\sigma\cdot\vec p|
\Lambda_{c1}^{+}(\frac12)\uparrow\rangle = -\sqrt{\frac23} I_S\\
& &\langle\Lambda_{c}^+\uparrow |\vec\sigma\cdot\vec p|
\Sigma_{c0}^{++}(\frac12)\uparrow\rangle = -\sqrt2 I_S\\
& &\langle\Sigma_{c}^+\uparrow |\vec\sigma\cdot\vec p|
\Sigma_{c1}^{++}(\frac12)\uparrow\rangle = -\frac{2}{\sqrt3} I_S\,,\label{31}
\eea
with $I_S$ an unknown reduced matrix element. Comparing these relations
with (\ref{24}-\ref{25}) gives immediately the first two relations
(\ref{23}).
The other relations in (\ref{23}) can be obtained in an analogous way
with the help of the quark model wavefunctions in Appendix.

Similar relations can be derived in the quark model among the 
D-wave amplitudes produced by the interaction Lagrangian (\ref{6D}), 
for which we obtain
\bea\label{DQM}
|h_8| = |h_9| = |h_{10}|\,,\qquad \frac{|h_{11}|}{|h_{10}|} = \sqrt2\,,\qquad
h_{12} = h_{13} = h_{14} = h_{15} = 0\,.
\eea
The derivation proceeds analogously as in the case of (\ref{23}).
The Noether axial current associated with the terms (\ref{6D}) in the
chiral Lagrangian is given by (up to terms containing pion fields)
\bea\label{NoetherD}
J_\mu^a &=& -ih_8\epsilon_{ijk}\left\{ 
\partial_\alpha [\bar S_\mu^{kl} t^a_{lj}R^i_\alpha ] + 
\partial_\alpha [\bar S_\alpha^{kl} t^a_{lj}R^i_\mu ] +
\frac23 v_\mu v\cdot\partial [\bar S_\alpha^{kl} t^a_{lj}R^i_\alpha ]\right\}\\
& & -\,ih_9\mbox{tr}\left\{
\partial_\alpha [\bar S_\mu t^a V_\alpha ] +
\partial_\alpha [\bar S_\alpha t^a V_\mu ] +
\frac23 v_\mu v\cdot\partial [\bar S_\alpha t^a V_\alpha ]\right\}\nonumber\\
& &-\, 2ih_{10}\epsilon_{ijk}\partial_\nu [\bar T_i t^a_{jl} X_{\mu\nu}^{kl}]
-h_{11}\left\{
\epsilon^{\alpha\mu\sigma\lambda}\partial_\nu \mbox{tr }
[\bar S_\alpha t^a X_{\nu\sigma}]v_\lambda +
\epsilon^{\alpha\nu\sigma\lambda}\partial_\nu \mbox{tr }
[\bar S_\alpha t^a X_{\mu\sigma}]v_\lambda
\right\}\nonumber\\
& & -\,ih_{12}\mbox{tr}\left\{
\partial_\alpha [\bar S_\mu t^a R'_\alpha ] +
\partial_\alpha [\bar S_\alpha t^a R'_\mu ] +
\frac23 v_\mu v\cdot\partial [\bar S_\alpha t^a R'_\alpha ]\right\}\nonumber\\
& & -ih_{13}\epsilon_{ijk}\left\{ 
\partial_\alpha [\bar S_\mu^{kl} t^a_{lj}V^{'i}_\alpha ] + 
\partial_\alpha [\bar S_\alpha^{kl} t^a_{lj}V^{'i}_\mu ] +
\frac23 v_\mu v\cdot\partial [\bar S_\alpha^{kl} t^a_{lj}V^{'i}_\alpha ]
\right\}\nonumber\\
& &-\, 2ih_{14}\partial_\nu [\bar T_i t^a_{ji} X_{\mu\nu}^{'j}]
-h_{15}\epsilon_{ijk}\left\{
\epsilon^{\alpha\mu\sigma\lambda}\partial_\nu 
[\bar S_\alpha^{kl} t^a_{lj} X_{\nu\sigma}^{'i}]v_\lambda +
\epsilon^{\alpha\nu\sigma\lambda}\partial_\nu
[\bar S_\alpha^{kl} t^a_{lj} X_{\mu\sigma}^{'i}]v_\lambda
\right\}\nonumber\,.
\eea
For this case it is the matrix elements of the space components of 
$J^a_\mu$
which are nonvanishing. In the quark model they can be expressed as
\bea\label{DCQM}
\langle f|\bar d\gamma^i\gamma_5 u|i\rangle \to
\langle f|\sigma^i e^{i\vec q\cdot\vec r}|i\rangle &=&
\langle f|\sigma^i|i\rangle  + 
\frac{i}{2}\langle f|\sigma^i (\vec q\cdot\vec r)
+ r^i  (\vec q\cdot\vec\sigma )
-\frac23 q^i(\vec\sigma\cdot\vec r)|i\rangle\\
& &\qquad + \mbox{ higher multipoles}\,,\nonumber
\eea
where $\vec q$ denotes the momentum of the current.
The first term vanishes as $i$ and $f$ have different orbital
angular momenta. The second one is nonvanishing and 
will produce the terms proportional to the D-wave couplings
in (\ref{NoetherD}). From (\ref{NoetherD}) we read off the
following expressions for a few typical matrix elements
\bea\label{D1}
& &\langle \Sigma_c^0(\frac12),+\frac12 |q^\mu J^{1-i2}_\mu|
\Lambda_{c1}^+(\frac32),
+\frac12\rangle = \frac{\sqrt2}{3}h_8 (q_1^2+q_2^2-2q_3^2)\\
& &\langle \Sigma_c^+(\frac12),+\frac12 |q^\mu J^{1-i2}_\mu|
\Sigma_{c1}^{++}(\frac32),
+\frac12\rangle = \frac13 h_9 (q_1^2+q_2^2-2q_3^2)\\
& &\langle \Lambda_c^+(\frac12),+\frac12 |q^\mu J^{1-i2}_\mu|
\Sigma_{c2}^{++}(\frac32),
+\frac12\rangle = -\frac{2}{\sqrt{15}} h_{10} (q_1^2+q_2^2-2q_3^2)\\
& &\langle \Sigma_c^+(\frac12),+\frac12 |q^\mu J^{1-i2}_\mu|
\Sigma_{c2}^{++}(\frac32),
+\frac12\rangle = \frac{1}{\sqrt{10}} h_{11} (q_1^2+q_2^2-2q_3^2)\,.
\label{D4}
\eea

The quark model counterpart of these matrix elements can be computed
by expressing them as (\ref{DCQM}) and using
the wavefunctions in the Appendix. We obtain
\bea\label{QMD1}
& &\langle \Sigma_c^0(\frac12),+\frac12 |q^\mu J^{1-i2}_\mu|
\Lambda_{c1}^+(\frac32),+\frac12\rangle =
 \frac13 I_D(q_1^2+q_2^2-2q_3^2)\\
& &\langle \Sigma_c^+(\frac12),+\frac12 |q^\mu J^{1-i2}_\mu|
\Sigma_{c1}^{++}(\frac32),
+\frac12\rangle = -\frac{1}{3\sqrt2} I_D(q_1^2+q_2^2-2q_3^2)\\
& &\langle \Lambda_c^+(\frac12),+\frac12 |q^\mu J^{1-i2}_\mu|
\Sigma_{c2}^{++}(\frac32),
+\frac12\rangle = -\frac{2}{\sqrt{30}}  I_D(q_1^2+q_2^2-2q_3^2)\\
& &\langle \Sigma_c^+(\frac12),+\frac12 |q^\mu J^{1-i2}_\mu|
\Sigma_{c2}^{++}(\frac32),
+\frac12\rangle = -\frac{1}{\sqrt{10}} I_D(q_1^2+q_2^2-2q_3^2)\,.
\label{QMD4}
\eea
$I_D$ is the reduced matrix element of the $J=1$ spherical tensor
$(r_-,\,r_3,\,r_+)$. Comparing (\ref{D1}-\ref{D4}) with
(\ref{QMD1}-\ref{QMD4}) gives the relations (\ref{DQM}) between the
D-wave couplings\footnote{Similar reductions in the number of
couplings in the quark model have been obtained in
\cite{FKT}.}.

The coupling constants of the symmetric p-wave baryons among themselves
appearing in the Lagrangian (\ref{Lsymm}) can be also
estimated in the constituent quark model. We obtain (in units of $g_A$)
\bea
f_1 &=& 0\,,\qquad f_2 = -1\,,\qquad f_3 = 2\,,\qquad
|f_4| = \sqrt{\frac23}\,,\\
|f_5| &=& 1\,,\qquad
|f_6| = \frac{1}{\sqrt2}\,,\qquad |f_7| = 2\sqrt{\frac23}\,,\qquad
|f_8| = \frac{1}{\sqrt2}\,.\nonumber
\eea

The values taken by the couplings of the antisymmetric p-wave states
(\ref{Lasymm}) in the constituent
quark model are (also in units of $g_A$)
\bea
f'_1 &=& 0\,,\qquad f'_2 = -\frac12\,,\qquad f'_3 = 1\,,\qquad
|f'_4| = \sqrt{\frac23}\,,\\
|f'_5| &=& 1\,,\qquad
|f'_6| = \frac{1}{\sqrt2}\,,\qquad |f'_7| = \sqrt{\frac23}\,,\qquad
|f'_8| = \frac{1}{2\sqrt2}\,.\nonumber
\eea

Furthermore, all the couplings of the symmetric to the antisymmetric
p-wave states vanish in the quark model.
This is a consequence of symmetry mismatch. For s-wave and symmetric
p-wave states, diquark states with symmetric (antisymmetric) flavor
wave functions must have corresponding symmetric (antisymmetric) spin
wave functions. This correlation is reversed for diquarks in the
antisymmetric p-states. These opposite symmetry correlations lead to
the vanishing of all coupling constants between the antisymmetric
p-states and the s-wave and symmetric p-wave baryons.

\section{Phenomenological applications}

The sum rules presented in this paper can be used to provide 
model-independent constraints on the pion couplings of the heavy
baryons which will be have to be satisfied by model computations
or phenomenological determinations. For example, the sum rule
(\ref{14}) gives the upper bound
\bea\label{g2bound}
g_2^2 \leq \frac23\,.
\eea
This inequality is satisfied by the SU(2) Skyrme model calculation
of \cite{GLM} who find
\bea
|g_2| = \frac{1}{\sqrt2}G_A - \frac{1}{3\sqrt2}g = 0.717-0.813
\eea
with $G_A=1.25$ the nucleon axial charge and $g$ the $BB^*\pi$
coupling. The numerical values shown correspond to the experimental 
bounds for $g$ given in \cite{Am}, $g^2 = 0.09-0.5$ (at 90\%
confidence limit).
The constituent quark model prediction (\ref{cqm}) \cite{yan}
for $g_2=0.612$ (with $g_A=0.75$)
satisfies also (\ref{g2bound}), saturating it in the limit
$g_A\to 1$.

One can also derive an upper bound on the
coupling $g_1$ from the AW sum rule (\ref{22}). It can be
made stronger by assuming that $g_2$ is known
\bea\label{g1bound}
g_1^2 \leq \frac83 - \frac43 g_2^2\,.
\eea
The calculation of \cite{GLM} gives
\bea
|g_1| = G_A - \frac13 g = 1.014-1.150
\eea
which satisfies (\ref{g1bound}). The constituent quark model
with $g_A=0.75$ predicts a somewhat smaller value
$|g_1|=1$ (\ref{cqm}).

The CLEO collaboration recently measured the masses and 
widths of the $\Sigma_c^{*++}$ and $\Sigma_c^{*0}$ baryons
\cite{CLEO3/2} (for an earlier measurement of the masses of
these states see \cite{SKAT}). They find
\bea\label{DelS*1}
& &\Delta_{\Sigma_c^{*++}} = M_{\Sigma_c^{*++}}- M_{\Lambda_c^+}
 = 234.5\pm 1.36  \mbox{ MeV}\,,\qquad
\Gamma(\Sigma_c^{*++}) = 17.9^{+5.5}_{-5.1} \mbox{ MeV}\\
& &\Delta_{\Sigma_c^{*0}} = M_{\Sigma_c^{*0}}- M_{\Lambda_c^+}
 = 232.6\pm 1.28 \mbox{ MeV}\,,\qquad
\Gamma(\Sigma_c^{*0}) = 13.0^{+5.45}_{-5.0} \mbox{ MeV}\,.\label{DelS*2}
\eea
The mass of $\Lambda_c^+$ is given by the Particle Data Group
\cite{PDG} to be
\bea
M_{\Lambda_c^+} = 2284.9\pm 0.6\mbox{ MeV}\,.
\eea
Neglecting the radiative decay width associated with the
decays $\Sigma_c^*\to\Sigma_c\gamma$, the total width of
these states is given by
\bea\label{Gamma*++}
& &\Gamma(\Sigma_c^{*++}) = \frac{g_2^2}{2\pi f_\pi^2}
\left(\frac{M_{\Lambda_c^+}}{M_{\Sigma_c^{*++}}}\right)
|\vec p_\pi\,|^3 = g_2^2(47.971^{+1.24}_{-1.22})  \mbox{ MeV}\\
& &\Gamma(\Sigma_c^{*0}) = \frac{g_2^2}{2\pi f_\pi^2}
\left(\frac{M_{\Lambda_c^+}}{M_{\Sigma_c^{*0}}}\right)\label{Gamma*0}
|\vec p_\pi\,|^3 = g_2^2(46.268^{+1.14}_{-1.13})  \mbox{ MeV}\,.
\eea
The CLEO data (\ref{DelS*1},\ref{DelS*2}) offer thus the possibility
of extracting
the coupling $g_2$. From (\ref{Gamma*++}) we obtain the
value $|g_2|=0.611^{+0.097}_{-0.101}$ and from (\ref{Gamma*0})
$|g_2|=0.530^{+0.109}_{-0.119}$. Averaging these two values we
obtain our final result
\bea\label{g2}
|g_2| = 0.570^{+0.137}_{-0.159}\,,
\eea
which (especially the one following from (\ref{Gamma*++}))
is in good agreement with the constituent quark model prediction
$|g_2|=0.612$. Similar determinations of $g_2$ have been presented
recently in \cite{Sav1,HYC}. Our result comes closer to the one
in \cite{HYC}.

One can use (\ref{g2}) to predict the widths of the $\Sigma_c$
baryons. The fit of \cite{PDG} gives the mass values
\bea\label{DelS}
& &\Delta_{S^{++}} = M_{\Sigma_c^{++}}- M_{\Lambda_c^+} = 167.95\pm 0.25 
\mbox{ MeV}\\
& &\Delta_{S^0} = M_{\Sigma_c^0}- M_{\Lambda_c^+} = 167.2\pm 0.4
\mbox{ MeV}\,.
\eea
The $\Sigma_c$ widths are
\bea\label{Gamma++}
\Gamma(\Sigma_c^{++}) &=& \frac{g_2^2}{2\pi f_\pi^2}
\left(\frac{M_{\Lambda_c^+}}{M_{\Sigma_c^{++}}}\right)
|\vec p_\pi\,|^3 = g_2^2(6.232^{+0.089}_{-0.088})\mbox{ MeV} = 
2.025^{+1.134}_{-0.987}\mbox{ MeV}\\
\Gamma(\Sigma_c^{0}) &=& \frac{g_2^2}{2\pi f_\pi^2}
\left(\frac{M_{\Lambda_c^+}}{M_{\Sigma_c^{0}}}\right)\label{Gamma0}
|\vec p_\pi\,|^3 = g_2^2(5.969^{+0.140}_{-0.139})\mbox{ MeV} = 
1.939^{+1.114}_{-0.954}\mbox{ MeV}\,.
\eea
These states are significantly narrower than their spin-3/2
counterparts, which explains why their widths have not been
yet measured.

We present also predictions for the I=1/2 spin-3/2 charmed
baryons $\Xi^{*0}_c$ and $\Xi^{*+}_c$. The latter has been
only recently discovered \cite{CLEOXi}. Their measured
parameters are
\bea\label{bound1}
M_{\Xi_c^{*0}} &=& M_{\Xi_c^{+}} + (178.2\pm 1.1)  \mbox{ MeV}\,,\qquad
\Gamma(\Xi_c^{*0}) < 5.5 \mbox{ MeV}\quad\cite{Avery}\\\label{bound2}
M_{\Xi_c^{*+}} &=& M_{\Xi_c^{0}} + (174.3\pm 1.1)  \mbox{ MeV}\,,\qquad
\Gamma(\Xi_c^{*+}) < 3.1 \mbox{ MeV}\quad\cite{CLEOXi}\,.
\eea
The masses of the ground state I=1/2 baryons are 
$M_{\Xi_c^{+}}=2465.6\pm 1.4$ MeV and $M_{\Xi_c^{0}}=2470.3\pm 1.8$ MeV
\cite{PDG}. We shall neglect the small admixture of SU(3) sextet into
the ground states as the corresponding mixing angle is small, of the order 
of a few degrees \cite{Ito,Fra,Sav1}. Including both the charged and
neutral pion channels we obtain
\bea
\Gamma(\Xi_c^{*0}) &=& g_2^2(7.712\pm 0.436)\mbox{ MeV} = 
1.230-4.074\mbox{ MeV}\\
\Gamma(\Xi_c^{*+}) &=& g_2^2(7.496\pm 0.446)\mbox{ MeV} = 
1.191-3.971\mbox{ MeV}\,,
\eea
which are consistent with the bounds (\ref{bound1}) and (\ref{bound2}).

It is interesting to note that the difference of couplings on 
the l.h.s. of the polarized AW sum rule
(\ref{1/2-3/2}) vanishes when the 
constituent quark model relations (\ref{cqm}) are used;
so does the r.h.s. of (\ref{1/2-3/2}) when the constituent
quark model relations (\ref{DQM}) are inserted in this relation.
If the quark model relations (\ref{DQM}) are not used,
the sum rule (\ref{1/2-3/2}) can be employed to give 
a model-independent proof of the
constituent quark model relations (\ref{cqm}) in the
large-$N_c$ limit. 
Recalling the scaling relations of the couplings
discussed in Sec.IV.B one can see that in the large-$N_c$
limit the r.h.s. is suppressed by one power of $1/N_c$
relative to the l.h.s.. 

It is, of course, a well-known fact that
the predictions of the constituent quark model for low-lying
s-wave baryon states become exact in the large-$N_c$ limit
\cite{DJM}. What is new here is that the relation (\ref{1/2-3/2})
gives also the corrections to this result, expressed in terms of
couplings of the higher states.

As a matter of fact, the sum rule (\ref{1/2-3/2}) suggests that
the quark model relations (\ref{cqm}) might work better than
one would expect from large-$N_c$ alone. First, only
D-wave couplings appear in (\ref{1/2-3/2}), whose contributions
are suppressed by factors of $\Delta^2/\Lambda_{\chi}^2$, with
$\Delta$ the excitation energies of the p-wave states and
$\Lambda_\chi\simeq 1$ GeV the chiral symmetry breaking scale.
An explicit calculation using the upper limit on $h_8$ determined 
below (\ref{h8}) shows that the contribution of the $h_8^2$ term 
on the r.h.s. of the sum rule (\ref{1/2-3/2}) is under 0.06.
Second, the alternating signs of the terms on the r.h.s.
of (\ref{1/2-3/2}) could enhance further the above mentioned
suppression.

We can expect therefore the quark model relation between
$g_1$ and $g_2$ to be valid at the order of 10\% or better.
We obtain in this way the following prediction for $g_1$
\bea
|g_1| = 2\sqrt{\frac23}|g_2| = 
0.931^{+0.224+0.093}_{-0.260-0.093}\,.
\eea
The possibility of determining $g_1$ in this way is particularly 
welcome as the decay $\Sigma_c^*\to\Sigma_c\pi$ 
is kinematically forbidden, making 
a direct extraction of $g_1$ impossible
(for an alternative method see \cite{HYC,Sav}).

In contrast to $g_1$ and $g_2$, the couplings of the p-wave
baryons are not completely predicted by the simple constituent
quark model.
To do so, it must be supplemented with additional dynamical
assumptions, which in turn will have to be used to determine the
wavefunction of the constituent quarks. The AW sum rules 
discussed in this
paper could be used to construct a model for the couplings of
the p-wave states.
Inserting the quark model relations (\ref{cqm}) and (\ref{23}) into the
sum rules (\ref{14}) and (\ref{22}) one common relation is obtained
\bea\label{AWcqm}
1 = g_A^2 + \frac38 h_4^2 + \frac43 h_{10}^2 |\vec p_\pi\,|^2 + \cdots\,.
\eea
This result demonstrates the consistency of the constituent quark
model with the AW sum rules (\ref{14}) and (\ref{22}).
Assuming saturation with the states shown in Table 1, this sum rule
determines all the couplings of the p-wave baryons up to an
additional free parameter, the ratio of the D-wave to S-wave couplings.
In principle this ratio could be fixed with experimental input.

We will use in the following a different approach, 
based on extracting the couplings $h_2$ and
$h_8$ of the lowest-lying p-wave baryons directly from
experimental data. These could be subsequently used, together
with the quark model relations (\ref{23}) and (\ref{DQM}), to
predict the couplings of all the other symmetric p-wave baryons.
The advantage of this approach consists in minimizing the
number of necessary assumptions, reducing the model dependence
to the application of the relations (\ref{23}) and (\ref{DQM}).

We will determine the allowed range of values for $h_2$ and
$h_8$ by using two different measurements. The first
is the CLEO  measurement of
the $\Lambda_{c_1}^+(2593)$ width \cite{LC1}
\bea\label{Lambdawidth}
\Gamma(\Lambda_{c_1}^+(2593)) = 3.9^{+2.4}_{-1.6} \mbox{ MeV}\,.
\eea

The state $\Lambda_{c_1}^+(2593)$ is the $J^P=1/2^-$ member
of the $s_\ell^{\pi_\ell}=1^-$ p-wave heavy quark doublet.
Its general properties have 
been discussed already by Cho \cite{cho2} (see also \cite{HDH}). 
It is known that it
decays predominantly in the two-pion mode, which is enhanced by
resonant effects due to the proximity of the $\Sigma_c$ pole.
These theoretical expectations have been confirmed by experiment
\cite{LC1,LC2}.

The decay rate for the process $\Lambda_{c_1}^+(2593)\to
\Lambda_c^+\pi^+\pi^-$ has been computed in \cite{cho2}.
To lowest order in chiral perturbation theory and in the heavy mass
limit it is given by 
\bea\label{rate1}
& &\frac{\mbox{d}\Gamma(\Lambda_{c1}^{+}(2593)\to
\Lambda_c^+\pi^+(E_1)\pi^-(E_2))} {\mbox{d}E_1 \mbox{d}E_2} =\\
& &\qquad
\frac{g_2^2}{16\pi^3 f_\pi^4}M_{\Lambda_c^+}\left\{
\vec p_2\,^2|A|^2 + \vec p_1\,^2|B|^2 + 2[E_1 E_2-p_1\cdot p_2]
\mbox{Re }(AB^*)\right\}\nonumber
\eea
with 
\bea
& &A(E_1,E_2) =\frac{h_2E_1}{\Delta_R-\Delta_{\Sigma_c^0}-E_1+
i\Gamma_{\Sigma_c^0}/2}-\frac{\frac23 h_8\vec p_1\,^2}
{\Delta_R-\Delta_{\Sigma_c^{*0}}-E_1+i\Gamma_{\Sigma_c^{*0}}/2}\\
& & \qquad+\frac{2h_8[E_1 E_2-p_1\cdot p_2]}
{\Delta_R-\Delta_{\Sigma_c^{*++}}-E_2+i\Gamma_{\Sigma_c^{*++}}/2}\,,
\nonumber\\
& &B(E_1,E_2;\Delta_{\Sigma_c^{(*)0}},\Delta_{\Sigma_c^{(*)++}}) =
A(E_2,E_1;\Delta_{\Sigma_c^{(*)++}},\Delta_{\Sigma_c^{(*)0}})\,.
\eea
The boundaries of the Dalitz plot for these decays are given by
\bea
& &(E_1)_{min} = m_\pi\,,\qquad (E_1)_{max} =
\frac{M_{\Lambda_{c1}^+}^2-2M_{\Lambda_c^+}m_\pi-M_{\Lambda_{c}^+}^2}
{2M_{\Lambda_{c1}^+}}\nonumber\\
& &(E_2)_{min,max} =
\frac{(E_1-M_{\Lambda_{c1}^+})(M_{\Lambda_{c1}^+}^2+2m_\pi^2
-2M_{\Lambda_{c1}^+}E_1-M_{\Lambda_{c}^+}^2)\pm
\sqrt{\Delta}}
{2(2M_{\Lambda_{c1}^+}E_1-M_{\Lambda_{c1}^+}^2-m_\pi^2)}
\eea
with
\bea
\Delta = (E_1^2-m_\pi^2)[(M_{\Lambda_{c1}^+}^2-2M_{\Lambda_{c1}^+}E_1
-M_{\Lambda_{c}^+}^2)^2-4M_{\Lambda_c^+}^2m_\pi^2]\,.
\eea
The boundaries of the Dalitz plots for the two-pion decays of the 
$\Lambda_{c1}^+$ states 
are shown in Fig.2. In the limit when the energy release $\Delta_R$
is much smaller than the mass of the decaying baryon $M_{\Lambda_{c1}^+}$
the phase space degenerates to the line $E_1+E_2=\Delta_R$
and the decay rate reduces to the expression originally
derived in \cite{cho2}
(we included here also the contribution of the
D-wave couplings which was neglected in \cite{cho2})
\bea\label{cho's1}
\frac{\mbox{d}\Gamma(\Lambda_{c1}^{+}(2593)\to
\Lambda_c^+\pi^+(E_1)\pi^-(E_2))} {\mbox{d}E_1} &=&
\frac{g_2^2}{8\pi^3f_\pi^4}\left(\frac{M_{\Lambda_c^+}}
{M_{\Lambda_{c_1}^{+}}}\right)
\sqrt{(E_1^2-m_\pi^2)(E_2^2-m_\pi^2)}\\
& &\hspace{-4cm}\times\, \left\{h_2^2\left(
\frac{E_1^2(E_2^2-m_\pi^2)}{(\Delta_R-\Delta_{\Sigma_c^{0}}-E_1)^2+
\frac14 \Gamma_{\Sigma_c^0}^2}+
\frac{E_2^2(E_1^2-m_\pi^2)}{(\Delta_R-\Delta_{\Sigma_c^{++}}-E_2)^2+
\frac14 \Gamma_{\Sigma_c^{++}}^2}
\right)\right.\nonumber\\
& &\left.\hspace{-7cm}+\frac89 h_8^2
(E_1^2-m_\pi^2)(E_2^2-m_\pi^2)\left(
\frac{E_2^2-m_\pi^2}{(\Delta_R-\Delta_{\Sigma_c^{*++}}-E_2)^2+
\frac14 \Gamma_{\Sigma_c^{*++}}^2}+
\frac{E_1^2-m_\pi^2}{(\Delta_R-\Delta_{\Sigma_c^{*0}}-E_1)^2+
\frac14 \Gamma_{\Sigma_c^{*0}}^2}
\right)\right\}\nonumber\,.
\eea
We denoted by $\Delta_i$ the excitation energies of the respective
states with respect to $\Lambda_c^+$. We will use in our
analysis below the value  \cite{PDG}
\bea\label{DelR}
& &\Delta_R = M_{\Lambda_{c_1}^{+}}- M_{\Lambda_c^+} = 308.6\pm 0.8 
\mbox{ MeV}\,.
\eea
The widths in the propagator denominators are given by
\bea
\Gamma_{\Sigma_c^{(*)}} = \frac{g_2^2}{2\pi 
f_\pi^2}\frac{M_{\Lambda_c^+}}
{M_{\Sigma_c^{(*)}}}|\vec p_\pi\,|^3\,.
\eea

The rate for $\Lambda_{c_1}^+(2593)\to\Lambda_c^+\pi^0\pi^0$
is given by formulas identical to (\ref{rate1},\ref{cho's1})
with an additional
factor of 1/2 due to the identity of the pions in the final state.
In addition, the substitutions 
\bea\label{subst}
\Delta_{\Sigma_c^{(*)++}},\,\Delta_{\Sigma_c^{(*)0}} \to
\Delta_{\Sigma_c^{(*)+}}
\eea
have to be made.
We will use in our calculations the following numerical values 
\bea
& &\Delta_{\Sigma_c^{+}} = M_{\Sigma_c^{+}}- M_{\Lambda_c^+}
 = 168.5\pm 0.7 \mbox{ MeV}\quad\cite{PDG}\\
& &\Delta_{\Sigma_c^{*+}} = M_{\Sigma_c^{*+}}- M_{\Lambda_c^+}
 = 233.5^{+2.4}_{-2.2} \mbox{ MeV}\,.
\eea
The state $\Sigma_c^{*+}$ has not yet been observed so we will
use for its mass the average of its two isospin partners 
(\ref{DelS*1},\ref{DelS*2}).

The second experimental input we will use is an upper bound on
the width of $\Lambda_{c1}^+(2625)$ obtained also by CLEO \cite{LC1}
\bea\label{width3/2}
\Gamma(\Lambda_{c_1}^+(2625)) < 1.9 \mbox{ MeV}\,.
\eea
$\Lambda_{c_1}^+(2625)$ is 
the spin-3/2 heavy quark symmetry partner of $\Lambda_{c_1}^+(2593)$.
To leading order in the heavy mass expansion, the two-pion decay rate
of this state is given by
\bea\label{rate2}
& &\frac{\mbox{d}\Gamma(\Lambda_{c1}^{+}(2625)\to
\Lambda_c^+\pi^+(E_1)\pi^-(E_2))} {\mbox{d}E_1 \mbox{d}E_2} =\\
& &\quad
\frac{g_2^2}{16\pi^3 f_\pi^4}M_{\Lambda_c^+}\left\{
\vec p_1\,^2|C|^2 + \vec p_2\,^2|E|^2 + 2[E_1 E_2-p_1\cdot p_2]
\mbox{Re }(CE^*) +
[\vec p_1\,^2\vec p_2\,^2-(E_1E_2-p_1\cdot p_2)^2]\right.\nonumber\\
& &\left.\quad\,\,\,\,\times
\left[ \vec p_1\,^2|D|^2 + \vec p_2\,^2|F|^2 -
\mbox{Re }(CF^*) + \mbox{Re }(DE^*) + 2(E_1E_2-p_1\cdot p_2)
\mbox{Re }(DF^*)\right]\right\}\nonumber
\eea
with
\bea
& &C(E_1,E_2) = \left(h_2E_2-\frac23 h_8\vec p_2\,^2\right)
\frac{1}{\Delta_{R^*}-\Delta_{\Sigma_c^{*++}}-E_2+
i\Gamma_{\Sigma_c^{*++}}/2}\\
& &\qquad+\,\frac23 h_8(E_1E_2-p_1\cdot p_2)
\left(\frac{1}{\Delta_{R^*}-\Delta_{\Sigma_c^{0}}-E_1+
i\Gamma_{\Sigma_c^{0}}/2} + 
\frac{2}{\Delta_{R^*}-\Delta_{\Sigma_c^{*0}}-E_1+
i\Gamma_{\Sigma_c^{*0}}/2}\right)\nonumber\\
& &D(E_1,E_2) = \frac23 h_8\left(
-\frac{1}{\Delta_{R^*}-\Delta_{\Sigma_c^{0}}-E_1+
i\Gamma_{\Sigma_c^{0}}/2} +
\frac{1}{\Delta_{R^*}-\Delta_{\Sigma_c^{*0}}-E_1+
i\Gamma_{\Sigma_c^{*0}}/2}\right)\\
& &E(E_1,E_2;\Delta_{\Sigma_c^{(*)0}},\Delta_{\Sigma_c^{(*)++}}) =
C(E_2,E_1;\Delta_{\Sigma_c^{(*)++}},\Delta_{\Sigma_c^{(*)0}})\,,\\
& &F(E_1,E_2;\Delta_{\Sigma_c^{(*)0}},\Delta_{\Sigma_c^{(*)++}}) =
-D(E_2,E_1;\Delta_{\Sigma_c^{(*)++}},\Delta_{\Sigma_c^{(*)0}})\,.
\eea
In analogy to the previous case,  this decay rate simplifies 
in the heavy mass
limit  and is given by
\bea\label{cho's2}
\frac{\mbox{d}\Gamma(\Lambda_{c_1}^{+}(2625)\to
\Lambda_c^+\pi^+(E_1)\pi^-(E_2))}
{\mbox{d}E_1} &=&
\frac{g_2^2}{8\pi^3f_\pi^4}\left(\frac{M_{\Lambda_c^+}}
{M_{\Lambda_{c_1}^{+}}}\right)
|\vec p_1\,|\,|\vec p_2\,|\\
& &\hspace{-4cm}\times\, \left\{h_2^2\left(
\frac{E_1^2\vec p_2\,^2}{(\Delta_{R^*}-\Delta_{\Sigma_c^{*0}}-E_1)^2+
\frac14 \Gamma_{\Sigma_c^{*0}}^2}+
\frac{E_2^2\vec p_1\,^2}{(\Delta_{R^*}-\Delta_{\Sigma_c^{*++}}-E_2)^2+
\frac14 \Gamma_{\Sigma_c^{*++}}^2}
\right)\right.\nonumber\\
& &\left.\hspace{-6cm}+\frac49 h_8^2
\vec p_1\,^2\vec p_2\,^2\left[\vec p_1\,^2\left(
\frac{1}{(\Delta_{R^*}-\Delta_{\Sigma_c^{0}}-E_1)^2+
\frac14 \Gamma_{\Sigma_c^0}^2}+
\frac{1}{(\Delta_{R^*}-\Delta_{\Sigma_c^{*0}}-E_1)^2+
\frac14 \Gamma_{\Sigma_c^{*0}}^2}\right)\right.\right.\nonumber\\
& &\left.\left.\hspace{-6cm}+\vec p_2\,^2\left(
\frac{1}{(\Delta_{R^*}-\Delta_{\Sigma_c^{++}}-E_2)^2+
\frac14 \Gamma_{\Sigma_c^{++}}^2}+
\frac{1}{(\Delta_{R^*}-\Delta_{\Sigma_c^{*++}}-E_2)^2+
\frac14 \Gamma_{\Sigma_c^{*++}}^2}\right)\right]\right\}\nonumber\,.
\eea
The rate for the neutral
pions channel can again be obtained by adding a symmetry factor 1/2
and making the substitutions (\ref{subst}) in this formula.
The excitation energy of this state is \cite{PDG} 
\bea
\Delta_{R^*} = M_{\Lambda_{c_1}^{+}(\frac32)} - M_{\Lambda_{c}^{+}} =
341.5\pm 0.8\mbox{ MeV}\,.
\eea

We will assume the widths of the two $\Lambda_{c1}^+$ baryons to
be dominated by their two-pion decay modes and neglect the
contribution of the multipion and radiative $\Lambda_{c1}^+\to
\Lambda_{c}^+\gamma$ modes. The latter approximation is supported
by model computations of the partial width for this mode 
\cite{CKRAD} which gave the small values 
$\Gamma(\Lambda_{c_1}^+(2593)\to \Lambda_c^+\gamma) = 0.016$ MeV and 
$\Gamma(\Lambda_{c_1}^+(2625)\to \Lambda_c^+\gamma) = 0.021$ MeV.

The total widths of the $\Lambda_{c1}^+$ states obtained by
integrating (\ref{rate1}) and (\ref{rate2}) (including also
the $\pi^0\pi^0$ channel)
 can be represented,
for given $g_2$, as two ellipses in the $(h_2,\,h_8)$ plane.
For example, the decay widths of the two $\Lambda_{c1}^+$ states
are given, for the central values of $g_2$ and hadron masses, by
\bea\label{G1/2}
\Gamma(\Lambda_{c1}^+(2593)) &=& 11.902h_2^2 + 13.817h_8^2 - 
0.042h_2 h_8\mbox{ (MeV)}\\\label{G3/2}
\Gamma(\Lambda_{c1}^+(2625)) &=& 0.518h_2^2 + (0.148\cdot 10^6)h_8^2 - 
5.229h_2 h_8\mbox{ (MeV)}\,.
\eea

The constraints on $(h_2,\,h_8)$ are plotted in Fig.3 for the 
interval of values for
$g_2$ (\ref{g2}). At the scale of the plot the two ellipses appear
very elongated, the vertical lines corresponding to the limits
on $\Gamma(\Lambda_{c_1}^+(2593))$ in (\ref{Lambdawidth}) and
the horizontal line giving an upper bound on $|h_8|$ arising from
(\ref{width3/2}). 
From Fig.3 we read off the following values
\bea\label{h2}
|h_2| &=& 0.572^{+0.322}_{-0.197}\\\label{h8}
|h_8| &\leq& (3.50-3.68)\,\cdot\,10^{-3}\mbox{ MeV}^{-1}\,.
\eea
The errors shown are mainly due to the uncertainties in the
masses (as the resonant decay $\Lambda_{c1}^+\to\Sigma_c\pi$ 
takes place very close to threshold) and in the total
widths of $\Lambda_{c1}^+(2593)$ and $\Lambda_{c1}^+(2625)$.
The error due to the unknown relative sign of $h_2$
and $h_8$ arising from the last terms in (\ref{G1/2},\ref{G3/2})
is negligible.
The upper bound on the ratio $h_8/h_2\leq 10^{-2}$ 
MeV$^{-1}$ is one order of magnitude above
the naive dimensional analysis estimate
$h_8/h_2 \simeq 1/\Lambda_\chi\simeq$ 10$^{-3}$ MeV$^{-1}$. 
The analogous couplings in the strange hyperons' system have been 
used in \cite{HDH} to obtain an estimate for the charm case.
Their results $|h_2|=0.54, |h_8|=0.55/\Lambda_\chi$ are
compatible with the values (\ref{h2},\ref{h8}) and suggest that
the upper bound (\ref{h8}) on $h_8$ 
overestimates the real value of this coupling by a factor of 10.
For illustration we quote also the values of the couplings
obtained when the simplified formulas (\ref{cho's1}) and
(\ref{cho's2}) are used instead of (\ref{rate1}) and (\ref{rate2}):
$h_2 = 0.553^{+0.310}_{-0.191}$, $|h_8|<3.63\cdot 10^{-3}$
MeV$^{-1}$.

Our results show that the width of the $\Lambda_{c1}^+(2625)$
state is possibly dominated by the D-wave term proportional to
$h_8^2$ which can become resonant. The S-wave contribution 
proportional to $h_2^2$ to the width accounts for 0.096-0.250 
MeV of the total.

In Fig.4 we show the allowed region for the couplings in the
$(g_2,\,h_2)$ plane, together with the constraints imposed by
the model-independent AW sum rule (\ref{+1/2})
\bea\label{modind}
1 \geq g_2^2 + \frac12 h_2^2 
\eea
and the constituent quark model version (\ref{AWcqm}) of the AW
sum rule
\bea\label{cqmsr}
1 \geq \frac32 g_2^2 + \frac32 h_2^2 \,.
\eea
We neglected the contribution of the D-wave couplings on the
r.h.s. of these sum rules.

The inequality (\ref{modind}) is satisfied for all values
of the $(g_2,\,h_2)$ parameters by the constraint on the width
of $\Lambda_{c1}^+(2593)$. On the other hand
the constituent quark sum rule (\ref{cqmsr}) is more 
restrictive. It favors smaller values for $h_2$:
$|h_2| = 0.375-0.705$.

We can use the extracted values for $h_2$ and $h_8$ (\ref{h2},\ref{h8})
to compute the resonant branching ratios of the $\Lambda_{c1}^+(2593)$
baryon. These quantities have been measured by the CLEO and E687
collaborations \cite{LC1,LC2,Luca}. The CLEO
results are
\bea\label{BRfraction1}
& &f_{\Sigma_c^{++}} =
\frac{BR(\Lambda_{c_1}^+(2593)\to\Sigma_c^{++}\pi^-)}
{BR(\Lambda_{c_1}^+(2593)\to\Lambda_c^+\pi^+\pi^-)} =
0.36\pm 0.13\\\label{BRfraction2}
& &f_{\Sigma_c^0} =
\frac{BR(\Lambda_{c_1}^+(2593)\to\Sigma_c^0\pi^-)}
{BR(\Lambda_{c_1}^+(2593)\to\Lambda_c^+\pi^+\pi^-)} =
0.42\pm 0.13
\eea
whereas the E687 collaboration \cite{LC2,Luca} only
quotes the total resonant branching ratio cummulated
over the type of $\Sigma_c$ 
\bea\label{BRfraction}
\frac{BR(\Lambda_{c_1}^+(2593)\to\Sigma_c\pi^\pm)}
{BR(\Lambda_{c_1}^+(2593)\to\Lambda_c^+\pi^+\pi^-)} &=&
0.90\pm 0.25\quad
(> 51\% \mbox{ at the 90\% confidence level})\,.
\eea

The experimental procedure for determining the resonant branching
fractions (\ref{BRfraction1},\ref{BRfraction2}) is based on measuring
the ratio
\bea\label{expBR}
f_{\Sigma_c} = \frac{BR(\Lambda_{c_1}^+(2593)\to\Sigma_c\pi^\pm)}
{BR(\Lambda_{c_1}^+(2593)\to\Lambda_c^+\pi^+\pi^-)} =
\frac{Y_{\Sigma_c\, region}-Y_{sideband}}{Y_{total}}\,.
\eea
Here $Y_{\Sigma_c\, region}$ is the number of events
for which the invariant mass of a pion and the $\Lambda_c^+$ is
within $\pm 4$ MeV from the mass of the corresponding $\Sigma_c$
state. $Y_{sideband}$ is the number of events for which
$E_1$ (the energy of the positively charged pion) is contained in
the sidebands  $(150.7,\, 155.1)$ MeV
for a $\Sigma_c^0$ and $(153.7,\, 158.1)$ MeV for a $\Sigma_c^{++}$
(for the E687 experiment \cite{LC2,Luca}).
These sidebands are introduced to eliminate the background
and have been chosen such that (\ref{expBR}) vanishes for a 
completely nonresonant process.

We do not know the sideband parameters for the CLEO experiment
\cite{LC1} so we will neglect $Y_{sideband}$ in (\ref{expBR}).
We obtain with the help of the theoretical expression 
(\ref{cho's1})
\bea\label{theor1}
f_{\Sigma_c^{++}} &=& 0.372^{+0.055}_{-0.074}\,,\qquad\qquad
f_{\Sigma_c^0} = 0.512^{+0.054}_{-0.061}
\eea
for the individual branching fractions.
For the cummulated branching fraction we use the E687 
sideband parameters
quoted above to obtain
\bea\label{theor2}
f_{\Sigma_c} &=& 0.860^{+0.085}_{-0.086}\,.
\eea
These values (except $f_{\Sigma_c^0}$ which is off by $1\sigma$) 
are in good agreement with the experimental data 
(\ref{BRfraction1}-\ref{BRfraction}).
The errors in (\ref{theor1}-\ref{theor2}) depend almost exclusively
on $g_2$ and the hadron masses and are insensitive to the 
precise value of $h_2$. This is due
to the negligibly small contribution made by the
term proportional to $h_8^2$ to the total rate of 
$\Lambda_{c1}^+(2593)$.

As mentioned previously, our results
(\ref{h2},\ref{h8}) can be used in conjunction with the quark model
relations (\ref{23},\ref{DQM}) to predict the couplings of all
symmetric p-wave baryons. Unfortunately, except for model
calculations \cite{CIK,CI}, the masses of these states are not yet 
known. Eventually they will be measured experimentally. 
For the time being we will limit ourselves to illustrating this 
application by using
the results of the model calculation \cite{CIK} for the baryon masses.

The next excitations above $\Lambda_{c1}^+$ are expected
to be the $\Sigma_{c0}(\frac12)$ baryons, which have the light 
degrees of freedom in a $s_\ell^{\pi_\ell}=0^-$ state.
Their excitation energy is estimated to be \cite{CIK}
\bea
\Delta_U = M_{\Sigma_{c0}^{++}} - M_{\Lambda_c^+} 
\simeq 500\mbox{ MeV}\,.
\eea
This state can decay directly to $\Lambda_c^+\pi^+$ in an S-wave
with a width
\bea
\Gamma(\Sigma_{c0}^{++}) = \frac{h_3^2}{2\pi f_\pi^2}
\frac{M_{\Lambda_c^+}}{M_{\Sigma_{c0}^{++}}} E_\pi^2|\vec p_\pi\,|\,.
\eea
Using the quark model relation $|h_3|=\sqrt3 |h_2|$ (\ref{23})
together with the value (\ref{h2}) for $h_2$ we 
obtain\footnote{This estimate for $h_3$ is compatible,
within its error bounds, with
the unitarity bound (\ref{SBound}) $|h_3|\leq 0.858$.}
\bea
\Gamma(\Sigma_{c0}^{++}) \simeq h_2^2 (2.066)\mbox{ GeV}
= 0.676 ^{+0.975}_{-0.385}\mbox{ GeV}\,.
\eea
This state is so broad that it might be very difficult to observe it.

The next symmetric baryons are $\Sigma_{c1}$, represented by
the superfield $V$ in (\ref{6}). Their dominant decay mode is
expected to be, as in the case of $\Lambda_{c1}^+$, into
two pions. The decay rate of the $\Sigma_{c1}^{++}(\frac12)$ is
\bea\label{rate3}
& &\frac{\mbox{d}\Gamma(\Sigma_{c1}^{++}(\frac12)\to
\Lambda_c^+\pi^+(E_1)\pi^0(E_2))} {\mbox{d}E_1 \mbox{d}E_2} =\\
& &\qquad
\frac{g_2^2}{32\pi^3 f_\pi^4}M_{\Lambda_c^+}\left\{
\vec p_2\,^2|A|^2 + \vec p_1\,^2|B|^2 + 2[E_1 E_2-p_1\cdot p_2]
\mbox{Re }(AB^*)\right\}\nonumber
\eea
with 
\bea
& &A(E_1,E_2) =\frac{h_4E_1}{\Delta_V-\Delta_{\Sigma_c^+}-E_1+
i\Gamma_{\Sigma_c^+}/2}-\frac{\frac23 h_9\vec p_1\,^2}
{\Delta_V-\Delta_{\Sigma_c^{*+}}-E_1+i\Gamma_{\Sigma_c^{*+}}/2}\\
& & \qquad-\,\frac{2h_9[E_1 E_2-p_1\cdot p_2]}
{\Delta_V-\Delta_{\Sigma_c^{*++}}-E_2+i\Gamma_{\Sigma_c^{*++}}/2}\,,
\nonumber\\
& &B(E_1,E_2;\Delta_{\Sigma_c^{(*)+}},\Delta_{\Sigma_c^{(*)++}}) =
-A(E_2,E_1;\Delta_{\Sigma_c^{(*)++}},\Delta_{\Sigma_c^{(*)+}})\,.
\eea
The decay rates for the isospin partners of $\Sigma_{c1}^{++}$ are
related as $\Gamma(\Sigma_{c1}^{(*)++,0}\to \Lambda_c^+\pi^0\pi^\pm) =
\Gamma(\Sigma_{c1}^{(*)+}\to \Lambda_c^+\pi^+\pi^-)$.
The decay $\Sigma_{c1}^{(*)+}\to \Lambda_c^+\pi^0\pi^0$ is
forbidden in the limit of exact isospin symmetry, due to the
fact that the two neutral pions cannot have isospin 1.

The decay rate of the $\Sigma_{c1}^{++}(\frac32)$ state is
given by a formula analogous to (\ref{rate2})
\bea\label{rate4}
& &\frac{\mbox{d}\Gamma(\Sigma_{c1}^{++}\to
\Lambda_c^+\pi^+(E_1)\pi^0(E_2))} {\mbox{d}E_1 \mbox{d}E_2} =\\
& &\quad
\frac{g_2^2}{32\pi^3 f_\pi^4}M_{\Lambda_c^+}\left\{
\vec p_1\,^2|C|^2 + \vec p_2\,^2|E|^2 + 2[E_1 E_2-p_1\cdot p_2]
\mbox{Re }(CE^*) +
[\vec p_1\,^2\vec p_2\,^2-(E_1E_2-p_1\cdot p_2)^2]\right.\nonumber\\
& &\left.\quad\,\,\,\,\times
\left[ \vec p_1\,^2|D|^2 + \vec p_2\,^2|F|^2 -
\mbox{Re }(CF^*) + \mbox{Re }(DE^*) + 2(E_1E_2-p_1\cdot p_2)
\mbox{Re }(DF^*)\right]\right\}\nonumber
\eea
with
\bea
& &C(E_1,E_2) = -\left(h_4E_2-\frac23 h_9\vec p_2\,^2\right)
\frac{1}{\Delta_{V^*}-\Delta_{\Sigma_c^{*++}}-E_2+
i\Gamma_{\Sigma_c^{*++}}/2}\\
& &\qquad+\,\frac23 h_9(E_1E_2-p_1\cdot p_2)
\left(\frac{1}{\Delta_{V^*}-\Delta_{\Sigma_c^{+}}-E_1+
i\Gamma_{\Sigma_c^{+}}/2} + 
\frac{2}{\Delta_{V^*}-\Delta_{\Sigma_c^{*+}}-E_1+
i\Gamma_{\Sigma_c^{*+}}/2}\right)\nonumber\\
& &D(E_1,E_2) = \frac23 h_9\left(
-\frac{1}{\Delta_{V^*}-\Delta_{\Sigma_c^{+}}-E_1+
i\Gamma_{\Sigma_c^{+}}/2} +
\frac{1}{\Delta_{V^*}-\Delta_{\Sigma_c^{*+}}-E_1+
i\Gamma_{\Sigma_c^{*+}}/2}\right)\\
& &E(E_1,E_2;\Delta_{\Sigma_c^{(*)+}},\Delta_{\Sigma_c^{(*)++}}) =
-C(E_2,E_1;\Delta_{\Sigma_c^{(*)++}},\Delta_{\Sigma_c^{(*)+}})\,,\\
& &F(E_1,E_2;\Delta_{\Sigma_c^{(*)+}},\Delta_{\Sigma_c^{(*)++}}) =
D(E_2,E_1;\Delta_{\Sigma_c^{(*)++}},\Delta_{\Sigma_c^{(*)+}})\,.
\eea

In addition to s-wave sextet baryons, the
$\Sigma_{c0}$ can appear also in the intermediate state.
The contribution of these diagrams can be expected to
be suppressed due to the large width of these states.
Neglecting them we obtain (for the central values of
$g_2$ and $\Sigma_c$ masses)
\bea
\Gamma(\Sigma_{c1}(\frac12)) &=& 81.3h_4^2 + (1.4\cdot 10^6)
h_9^2 - 33.4h_4h_9\mbox{  (MeV)} = 106.4^{+153.5}_{-60.7}\mbox{  MeV}\\
\Gamma(\Sigma_{c1}(\frac32)) &=& 72.3h_4^2 + (2.5\cdot
10^6) h_9^2 + 617.7h_4h_9\mbox{  (MeV)} = 94.7^{+136.6}_{-54.0}
\mbox{  MeV}
\eea
for $\Delta_{V^{(*)}}=465.1$ MeV (corresponding to $M_{\Sigma_{c1}}=2750$ 
MeV) and
\bea
\Gamma(\Sigma_{c1}(\frac12)) &=& 65.5h_4^2 + (5.1\cdot 10^6)
h_9^2 - 87.1h_4h_9\mbox{  (MeV)} = 85.8^{+123.7}_{-48.9}\mbox{  MeV}\\
\Gamma(\Sigma_{c1}(\frac32)) &=& 132.5h_4^2 + (4.7\cdot
10^6) h_9^2 + 1765.2h_4h_9\mbox{  (MeV)} = 173.4^{+250.1}_{-98.8}
\mbox{  MeV}
\eea
for $\Delta_{V^{(*)}}=515.1$ MeV (corresponding to $M_{\Sigma_{c1}}=2800$ 
MeV) respectively \cite{CIK}.
In the last equality the quark
model relations $|h_4|=2|h_2|$ and $|h_9|=|h_8|$ have been used
together with (\ref{h2},\ref{h8}). The terms proportional to $h_9$
have been neglected, as they are of the order of a few MeV.
These states appear to be considerably
broader than $\Lambda_{c1}^+$ due to the larger available phase
space. 

On the other hand, the $\Sigma_{c2}$ baryons have only
D-wave couplings. Their dominant decay mode is expected
to be two-body decay to $\Lambda_c^+\pi^+$ 
\bea
\Gamma(\Sigma_{c2}^{++}(\frac32,\frac52)\to\Lambda_c^+\pi^+) =
\frac{4h_{10}^2}{15\pi f_\pi^2}
\frac{M_{\Lambda_c^+}}{M_{\Sigma_{c2}^{++}}}
|\vec p_\pi\,|^5 \simeq 12\mbox{ MeV}\,,
\eea
where we used $M_{\Sigma_{c2}}=2800$ MeV \cite{CIK} and
the naive dimensional analysis estimate $|h_{10}|=0.4\cdot 10^{-3}$ 
MeV$^{-1}$.
In addition to this mode, the $\Sigma_{c2}$ baryons can also
decay to $\Sigma_c^{(*)}\pi$. The corresponding partial widths are
\bea
& &\Gamma(\Sigma_{c2}^{++}(\frac32)\to\Sigma_c^+\pi^+) +
\Gamma(\Sigma_{c2}^{++}(\frac32)\to\Sigma_c^{*+}\pi^+) =
\frac{h_{11}^2}{10\pi f_\pi^2}
\frac{M_{\Sigma_c^+}}{M_{\Sigma_{c2}^{++}}}
|\vec p_\pi\,|^5 +
\frac{h_{11}^2}{10\pi f_\pi^2}
\frac{M_{\Sigma_c^{*+}}}{M_{\Sigma_{c2}^{++}}}
|\vec p_\pi\,|^5\\
& &\Gamma(\Sigma_{c2}^{++}(\frac52)\to\Sigma_c^+\pi^+) +
\Gamma(\Sigma_{c2}^{++}(\frac52)\to\Sigma_c^{*+}\pi^+) =
\frac{2h_{11}^2}{45\pi f_\pi^2}
\frac{M_{\Sigma_c^+}}{M_{\Sigma_{c2}^{++}}}
|\vec p_\pi\,|^5 +
\frac{7h_{11}^2}{45\pi f_\pi^2}
\frac{M_{\Sigma_c^{*+}}}{M_{\Sigma_{c2}^{++}}}
|\vec p_\pi\,|^5
\eea
and identical formulas for the $\Sigma_c^{(*)++}\pi^0$ final states.
Adding together the contributions of all possible final states 
we obtain
\bea
\Gamma(\Sigma_{c2}^{++}(\frac32)\to\Sigma_c^{(*)}\pi) =
(9.86\cdot 10^{-6})h_{11}^2 \simeq 3.16\mbox{ MeV}\\
\Gamma(\Sigma_{c2}^{++}(\frac52)\to\Sigma_c^{(*)}\pi) =
(6.88\cdot 10^{-6})h_{11}^2 \simeq 2.20\mbox{ MeV}\,,
\eea
where we used the quark model relation $h_{11}^2=2h_{10}^2$
and the above-mentioned dimensional analysis estimate for
$h_{10}$. It must be mentioned  that
even a small mixing of $\Sigma_{c2}(\frac32)$ with 
the broader $\Sigma_{c1}(\frac32)$ could enhance its decay width.

The most surprising prediction concerns the remarkable stability of
the antisymmetric p-wave states $\Sigma'_{c1},\Lambda'_{c0},\Lambda'_{c1},
\Lambda'_{c2}$. The quark model predicts that these states do not decay
directly to s-wave baryons (see (\ref{23})), nor can they decay through
pole-mediated processes, with a symmetric p-wave state in the intermediate
state, as the two states do not couple to each other.

The corrections to this result come from including in the
quark model terms describing the pion coupling to two constituent
quarks simultaneously. Although these terms are formally of
order $1/N_c$ compared to the one-quark terms, their suppression
can be overcome as they can act coherently on the low-spin states
\cite{Wi,CGKM}. However, a fit performed to the strong decay
amplitudes of the light p-wave baryons \cite{CGKM} showed that,
contrary to the naive large-$N_c$ arguments, the two-quark operators 
turn out to be relatively unimportant. Presumably a similar situation
will hold true for the heavy p-wave baryons, in which case
neglecting these terms can be considered to be a good approximation.

Additional corrections to these results can be expected  from the
fact that in general, states with identical quantum numbers can mix. 
This could be an
important effect with the close pairs of states $(\Sigma_{c1}(\frac12)\,,
\Sigma'_{c1}(\frac12))$ and $(\Sigma_{c1}(\frac32)\,,
\Sigma'_{c1}(\frac32))$, which can mix even in the
heavy mass limit. On the other hand, mixing between
states belonging to heavy quark doublets with different values
for the quantum numbers of the light degrees of freedom $s_\ell^{\pi_\ell}$
is a $1/m_Q$ effect. Still, experimental evidence of mixing in the
system of charmed p-wave mesons \cite{PDG} shows that such $1/m_Q$ effects 
can be significant in the case of the charm heavy quark.

Another possible decay mode for the antisymmetric p-wave baryons is to the
channels $[ND]$ and $[ND^*]$. The threshold for the first one is at
2810 MeV and for the second one at 2950 MeV. Quark model calculations 
\cite{CIK,CI}
suggest that the p-wave baryons must be lighter than 3 GeV with some of the
states lying above these thresholds such that these
modes may well turn out to be significant. Unfortunately, at the present time
it is not possible
to treat these processes in a chiral perturbation theory framework, as done
in the pion decay case. There are, nevertheless, a few model-independent
predictions which can be made about these decays, following \cite{IW}.

The dominant decays can be expected to be (if kinematically allowed)
\bea
& &\Lambda'_{c0}(\frac12) \to [ND]_S\,,\quad [ND^*]_S\\
& &\Sigma'_{c1}(\frac12)\,,\Lambda'_{c1}(\frac12) \to [ND]_S\,,\quad [ND^*]_S\\
& &\Sigma'_{c1}(\frac32)\,,\Lambda'_{c1}(\frac32) \to [ND^*]_S\,,
\eea
which can proceed by S-waves. The decay $\Lambda'_{c2}(\frac32)\to [ND^*]_S$, 
although allowed by angular momentum and parity conservation, is
forbidden in the heavy mass limit.

Heavy quark symmetry predicts the following typical decay rate ratios
\bea\label{IWratio1}
& &\Gamma(\Lambda'_{c0}(\frac12) \to [ND]_S) : 
\Gamma(\Lambda'_{c0}(\frac12) \to [ND^*]_S) = 1\,:\,3\\
& &\Gamma(\Lambda'_{c1}(\frac12) \to [ND]_S) :
\Gamma(\Lambda'_{c1}(\frac12) \to [ND^*]_S) : 
\Gamma(\Lambda'_{c1}(\frac32) \to [ND]_S) : 
\Gamma(\Lambda'_{c1}(\frac32) \to [ND^*]_S) \nonumber\\
& &\qquad\qquad =\, \frac34\,:\,
\frac14\,:\,0\,:\,1\\
& &\Gamma(\Lambda'_{c1}(\frac12) \to [ND]_D) :
\Gamma(\Lambda'_{c1}(\frac12) \to [ND^*]_D) : 
\Gamma(\Lambda'_{c1}(\frac32) \to [ND]_D) : 
\Gamma(\Lambda'_{c1}(\frac32) \to [ND^*]_D) \nonumber\\
& &\qquad\qquad =\, 0\,:\,
1\,:\,\frac38\,:\,\frac58\\
& &\Gamma(\Lambda'_{c2}(\frac32) \to [ND]_D) : 
\Gamma(\Lambda'_{c2}(\frac32) \to [ND^*]_D) : 
\Gamma(\Lambda'_{c2}(\frac52) \to [ND]_D) : 
\Gamma(\Lambda'_{c2}(\frac52) \to [ND^*]_D) \nonumber\\
& &\qquad\qquad =\, \frac{5}{8}\,:\,
\frac{11}{8}\,:\,\frac{5}{12}\,:\,\frac{19}{12}\,.\label{IWratio2}
\eea
Kinematical effects such as mass splittings within the heavy quark symmetry
doublets and mixings among different states will certainly modify
these results. After accounting for these corrections,
the width ratios (\ref{IWratio1}-\ref{IWratio2}) can be expected to
be useful in identifying the heavy quark symmetry assignments
of these states.

\section{Conclusions}

In this paper we have made a systematic study of the strong interactions
of the s- and p-wave baryons containing a heavy quark. The dynamics of
these baryons are very rich. The richness is reflected by the large
number of multiplets (2 for s-wave and 8 for p-wave), and the large
number of coupling constants necessary to describe all the
interactions. We have found that the constituent quark model in
conjunction with the Adler-Weisberger sum rules provides a powerful
tool to handle the system. The quark model reduces the number of
coupling constants from 46 to 3 which are further constrained by a AW
sum rule. One of the three parameters is the axial vector coupling 
$g_A$ for the single quark transition $u\to d$. If we assume that $g_A$ 
is independent of the light quark environment, then its value is known to
be $g_A=0.75$ from the nucleon beta decay. The recent data on charmed
baryons from Fermilab and CLEO are consistent with this value of $g_A$ and
give strong constraints on the other two unknowns as discussed in 
Section VI.

Through the common value of $g_A$ the single coupling constant which is
needed to describe the s-wave heavy mesons is related to many of the
coupling constants in the heavy baryons. The choice of $g_A=0.75$ gives a
satisfactory rendition of the branching ratios of D$^*$ \cite{cheng}
and the decay widths of the charmed baryons as we
have seen in the last Section. However, this value of $g_A$ implies a
value for the DD$^*\pi$ coupling constant $g$ to be order of 0.7 \cite{HYC}
which is much larger than the values around 0.3 obtained by
other approaches such as QCD sum rules (see the references cited in
\cite{HYC}). It is therefore of great importance to measure the
width of D$^*$ to give a direct measurement of $g$. It will also confirm or
reject the hypothesis of environmental independence of $g_A$.

By heavy quark symmetry the bottom baryons are described by the same
interactions and the same coupling constants as those studied in this paper. 
In fact, heavy quark symmetry should work even better for the heavier
bottom baryons. With more forthcoming data on charmed baryons from
Fermilab (FOCUS) and CLEO and a wealth of data on bottom baryons expected 
from LEP and the B factories under construction in a few years, we hope that
many of our predictions will be tested experimentally in the near
future.

\acknowledgements

We thank Chi-Keung Chow for participating in the initial phase of
this work and for suggesting the application to $[ND]$ decays 
discussed in Section VI and Hai-Yang Cheng for useful comments on the
manuscript. We are grateful to Luca Cinquini and John Yelton for 
enlightening discussions about the E687 and  CLEO 
experimental results. D.P. is indebted to J\"urgen K\"orner for informing
him about \cite{FKT} before publication. The research of D.P. was supported
by the Ministry of Science and the Arts of Israel.
The work of T.M.Y. was supported in part by the National Science 
Foundation.

\newpage
\appendix
\section{Quark model wavefunctions for heavy baryons}

{\em s-wave}

\bea\label{A1}
& &|\Lambda_{c}^+\uparrow\rangle = |c\uparrow\rangle
\frac{1}{\sqrt2}(|\uparrow\downarrow\rangle - |\downarrow\uparrow\rangle)
\frac{1}{\sqrt2}(|ud\rangle - |du\rangle)\\\label{A2}
& &|\Sigma_{c}^+\uparrow\rangle =
\left(\sqrt{\frac23}|c\downarrow\rangle |\uparrow \uparrow\rangle -
\frac{1}{\sqrt6}|c\uparrow\rangle (|\uparrow \downarrow\rangle +
|\downarrow\uparrow\rangle)\right) \frac{1}{\sqrt2}(|ud\rangle+|du\rangle)
\eea

For the p-wave baryons our phase convention corresponds to combining
the total spin $S=s_1+s_2$ with the orbital momentum $L$ in the
order $S\otimes L$.

{\em p-wave (symmetric)}

\bea\label{A3}
& &|\Lambda_{c1}^+(\frac12),+\frac12\rangle =
(\sqrt{\frac23}|L(+1)c\downarrow\rangle - \frac{1}{\sqrt3}|L(0)c\uparrow\rangle)
\frac{1}{\sqrt2}(|\uparrow\downarrow\rangle - |\downarrow\uparrow\rangle)
\frac{1}{\sqrt2}(|ud\rangle - |du\rangle)\\
& &|\Lambda_{c1}^+(\frac32),+\frac12\rangle =
(\frac{1}{\sqrt3}|c\downarrow L(+1)\rangle + \sqrt{\frac23}|c\uparrow L(0)\rangle)
\frac{1}{\sqrt2}(|\uparrow\downarrow\rangle - |\downarrow\uparrow\rangle)
\frac{1}{\sqrt2}(|ud\rangle - |du\rangle)\\
& &|\Sigma_{c0}^{++}(\frac12),+\frac12\rangle =
\frac{1}{\sqrt3}|c\uparrow\rangle \left( 
|L(+1)u\downarrow u\downarrow\rangle - \frac{1}{\sqrt{2}}
|L(0)u\uparrow u\downarrow\rangle\right.\\
& &\left.\qquad\qquad -\,\frac{1}{\sqrt{2}}
|L(0)u\downarrow u\uparrow\rangle +
|L(-1)u\uparrow u\uparrow\rangle \right)\nonumber\\\label{A4}
& &|\Sigma_{c1}^{++}(\frac12),+\frac12\rangle =
-\frac{1}{\sqrt6}
|c\downarrow L(+1)\rangle
\left( |u\uparrow u\downarrow\rangle +
|u\downarrow u\uparrow\rangle \right) +
\frac{1}{\sqrt3}|c\downarrow L(0)u\uparrow u\uparrow\rangle\\
& &\qquad\qquad+\,\frac{1}{\sqrt6}
|c\uparrow L(+1)u\downarrow u\downarrow\rangle -
\frac{1}{\sqrt6}
|c\uparrow L(-1)u\uparrow u\uparrow\rangle\nonumber \\
& &|\Sigma_{c1}^{++}(\frac32),+\frac12\rangle =
\frac{1}{\sqrt6}
|c\downarrow \rangle
\left( |L(0)u\uparrow u\uparrow\rangle - \frac{1}{\sqrt2}
|L(+1)\rangle ( |u\uparrow u\downarrow\rangle +
|u\downarrow u\uparrow\rangle )\right) \\
& &\qquad\qquad +
\frac{1}{\sqrt3}|c\uparrow\rangle \left( |L(-1)u\uparrow u\uparrow\rangle
- |L(+1)u\downarrow u\downarrow\rangle \right)\nonumber \\
& &|\Sigma_{c2}^{++}(\frac32),+\frac12\rangle =
|c\downarrow \rangle
\left(\sqrt{\frac{3}{10}} |L(0)u\uparrow u\uparrow\rangle + \sqrt{\frac{3}{20}}
|L(+1)\rangle ( |u\uparrow u\downarrow\rangle + |u\downarrow u\uparrow\rangle )\right) \\
& &\qquad\qquad -
\frac{1}{\sqrt{15}}|c\uparrow\rangle
\left( |L(-1)u\uparrow u\uparrow\rangle
+ \sqrt2 |L(0)\rangle (|u\uparrow u\downarrow\rangle + |u\downarrow u\uparrow\rangle)
+ |L(+1)u\downarrow u\downarrow\rangle \right)\nonumber
\eea

{\em p-wave (antisymmetric)}

\bea\label{}
& &|\Sigma_{c1}^{'++}(\frac12),+\frac12\rangle =
\left(\sqrt{\frac23}|L(+1)c\downarrow\rangle - 
\frac{1}{\sqrt3}|L(0)c\uparrow\rangle\right)\cdot
\frac{1}{\sqrt2}(|u\uparrow u\downarrow\rangle-|u\downarrow u\uparrow\rangle)\\
& &|\Sigma_{c1}^{'++}(\frac32),+\frac12\rangle =
\left(\sqrt{\frac13}|L(+1)c\downarrow\rangle + 
\sqrt{\frac23}|L(0)c\uparrow\rangle\right)\cdot
\frac{1}{\sqrt2}(|u\uparrow u\downarrow\rangle-|u\downarrow u\uparrow\rangle)\\
& &|\Xi_{c0}^{'+}(\frac12),+\frac12\rangle =
\frac{1}{\sqrt6}|c\uparrow\rangle\left\{
|L(-1)\rangle (|u\uparrow s\uparrow\rangle-|s\uparrow u\uparrow\rangle)
\right.\nonumber\\
& &\qquad\qquad\qquad\left.-\frac{1}{\sqrt2}|L(0)\rangle
(|u\uparrow s\downarrow\rangle-|s\uparrow u\downarrow\rangle+
|u\downarrow s\uparrow\rangle - |s\downarrow u\uparrow\rangle)\right.\nonumber\\
& &\qquad\qquad\qquad\left.+\, |L(+1)\rangle
(|u\downarrow s\downarrow\rangle-|s\downarrow u\downarrow\rangle)\right\}\\
& &|\Lambda_{c1}^{'+}(\frac12),+\frac12 \rangle =
\frac{1}{\sqrt3}|c\downarrow\rangle\left( |L(0)\uparrow\uparrow\rangle
-\frac{1}{\sqrt2}|L(+1)\rangle (|\uparrow\downarrow\rangle+|\downarrow\uparrow\rangle)
\right)\frac{1}{\sqrt2}(|ud\rangle-|du\rangle)\nonumber\\
& &\qquad\qquad\qquad-\frac{1}{\sqrt6}|c\uparrow\rangle
\left(|L(-1)\uparrow\uparrow\rangle - |L(+1)\downarrow\downarrow\rangle\right)
\frac{1}{\sqrt2}(|ud\rangle-|du\rangle)\\
& &|\Lambda_{c1}^{'+}(\frac32),+\frac12 \rangle =
\frac{1}{\sqrt6}|c\downarrow\rangle\left( |L(0)\uparrow\uparrow\rangle
-\frac{1}{\sqrt2}|L(+1)\rangle (|\uparrow\downarrow\rangle+|\downarrow\uparrow\rangle)
\right)\frac{1}{\sqrt2}(|ud\rangle-|du\rangle)\nonumber\\
& &\qquad\qquad\qquad+\frac{1}{\sqrt3}|c\uparrow\rangle
\left(|L(-1)\uparrow\uparrow\rangle - |L(+1)\downarrow\downarrow\rangle\right)
\frac{1}{\sqrt2}(|ud\rangle-|du\rangle)\\
& &|\Xi_{c2}^{'+}(\frac32),+\frac12 \rangle =
\sqrt{\frac{3}{10}}|c\downarrow\rangle\left( |L(0)\uparrow\uparrow\rangle
+\frac{1}{\sqrt2}|L(+1)\rangle (|\uparrow\downarrow\rangle+|\downarrow\uparrow\rangle)
\right)\frac{1}{\sqrt2}(|us\rangle-|su\rangle)\nonumber\\
& &\qquad\qquad\qquad-\frac{1}{\sqrt{15}}|c\uparrow\rangle
\left(|L(-1)\uparrow\uparrow\rangle + \sqrt2 |L(0)\rangle 
(|\uparrow\downarrow\rangle + |\downarrow\uparrow\rangle)
+ |L(+1)\rangle |\downarrow\downarrow\rangle\right)\\
& &\qquad\qquad\qquad\times \frac{1}{\sqrt2}(|us\rangle-|su\rangle)\,.\nonumber
\eea

\newpage
\noindent
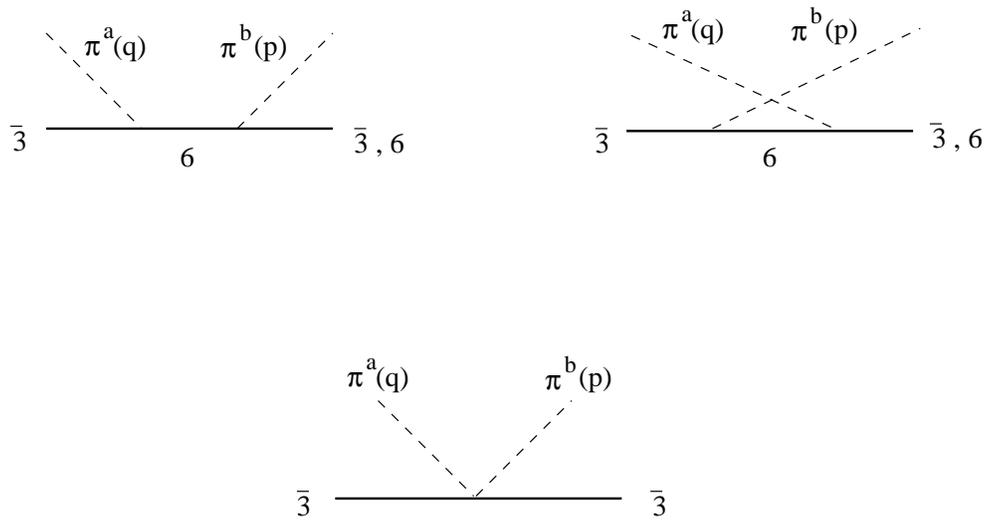
{\bf Fig.1} Feynman diagrams for the continuum contribution to the AW sum rule
on a ${\bf\bar 3}$ baryon.\\[0.5cm]

\noindent
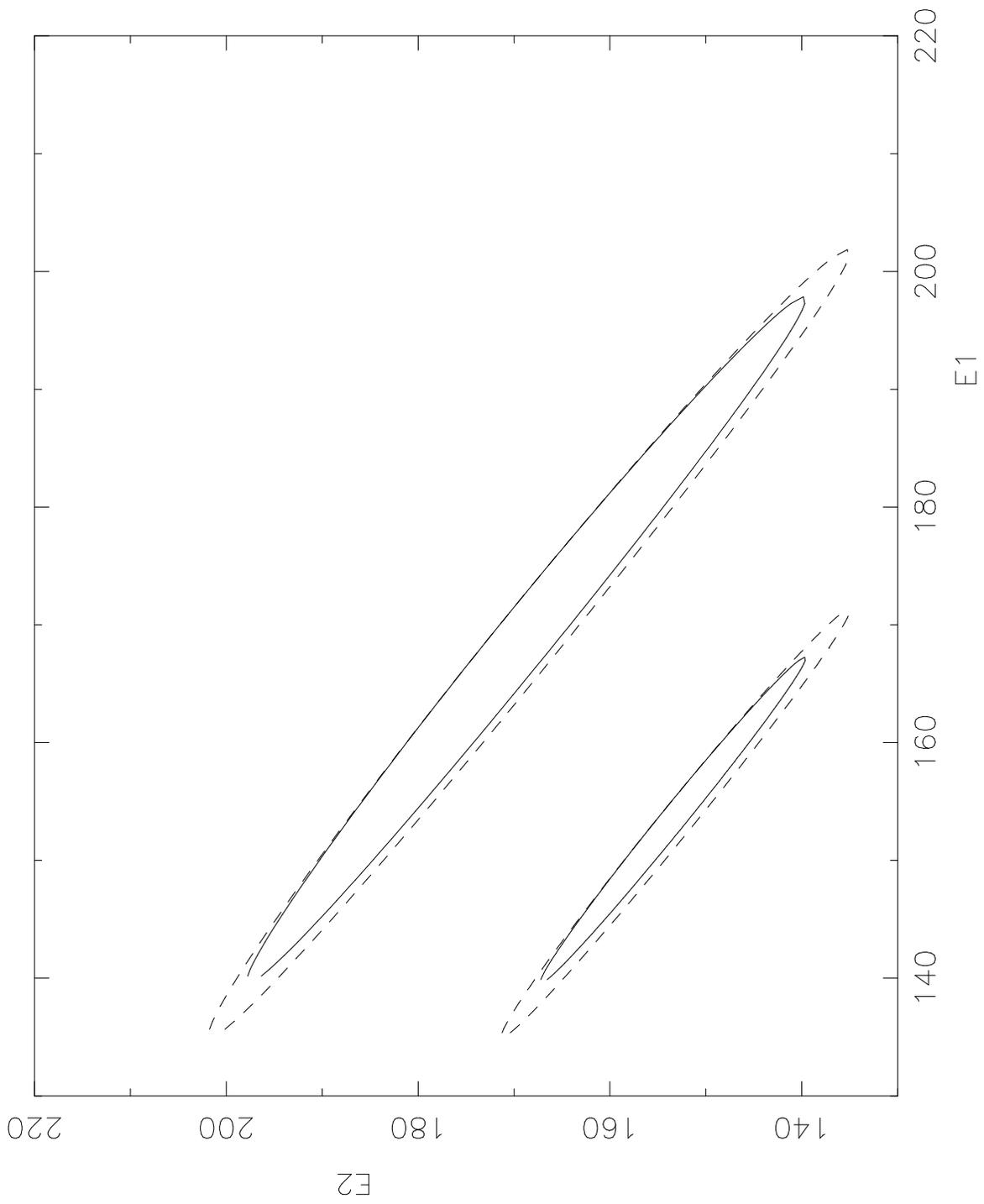
{\bf Fig.2} Boundaries of the Dalitz plots for the decays 
$\Lambda_{c1}^+(2593)\to
\Lambda_{c1}^+\pi\pi$ (lower) and $\Lambda_{c1}^+(2625)\to
\Lambda_{c1}^+\pi\pi$ (upper). The continuous lines correspond
to charged pions and dashed lines to neutral pions in the final
state respectively.\\[0.5cm]

\noindent
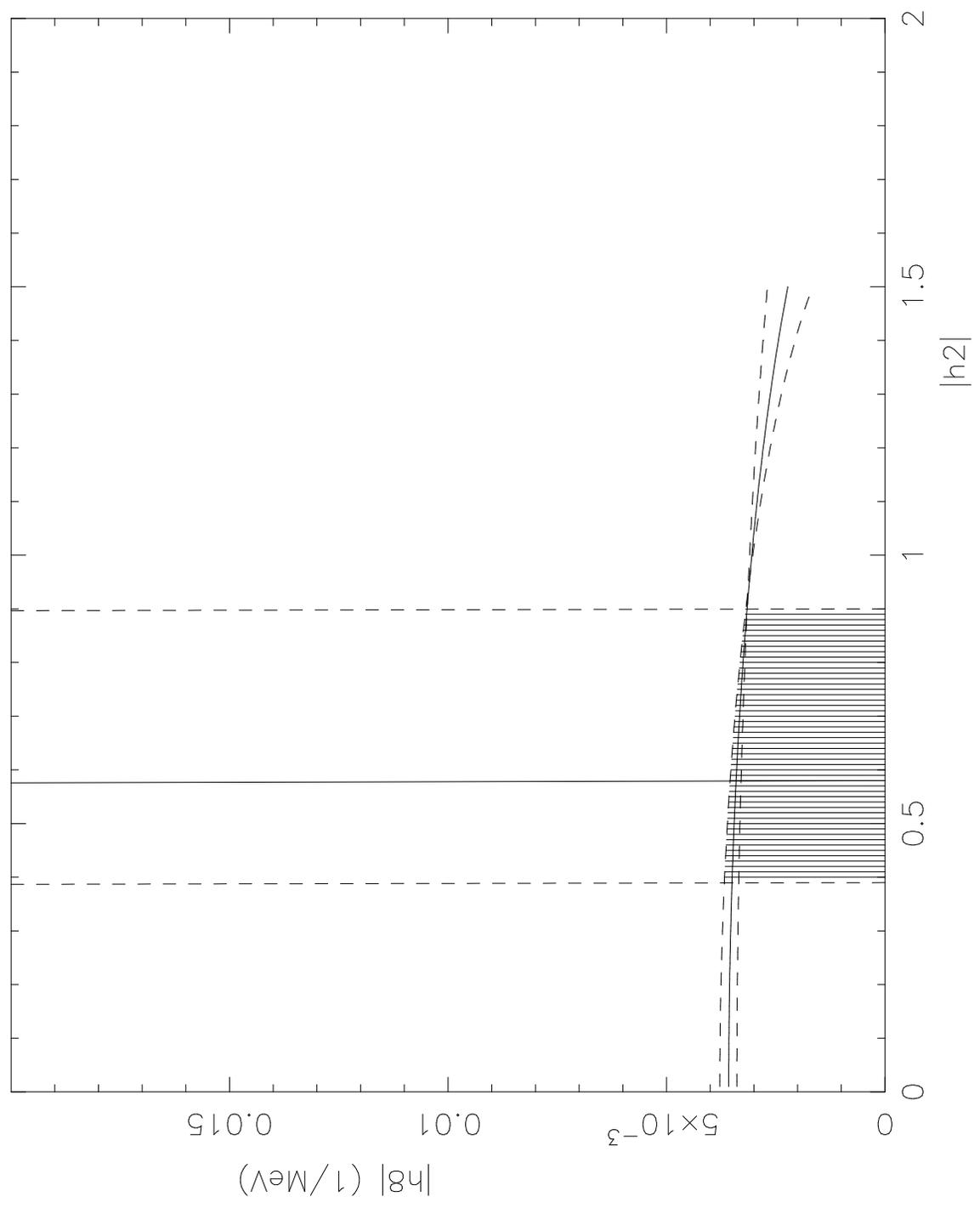
{\bf Fig.3} Constraints on the pion couplings of the $\Lambda_{c1}^+$
p-wave baryons $h_2$, $h_8$ from data on their decay widths.
The continuous lines give central values and the dashed lines
show 1$\sigma$ deviations.\\[0.5cm]

\noindent
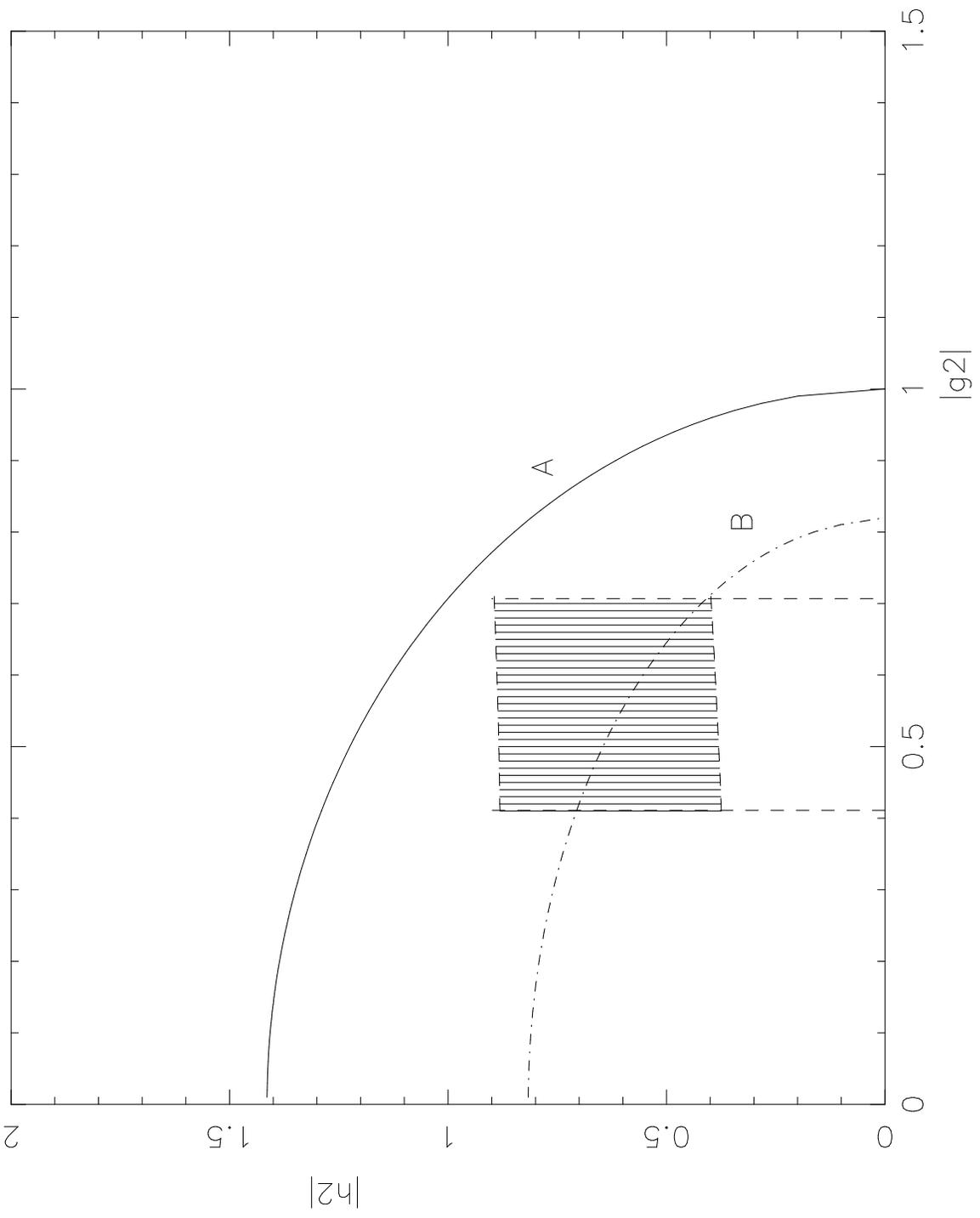
{\bf Fig.4} Allowed region for the couplings $(g_2,\, h_2)$
from the decay width of $\Lambda_{c1}(2593)$ and the
Adler-Weisberger sum rule (6.30) (curve A). The dash-dotted line B
shows the constraint imposed by the constituent quark model
AW sum rule (6.31). 

\end{document}